\documentclass[12pt,a4paper]{article}

\usepackage[british]{babel}

\usepackage[margin=1in]{geometry}
\usepackage{lineno}
\usepackage{setspace}
\usepackage[authoryear]{natbib}
\PassOptionsToPackage{hyphens}{url}
\usepackage{hyperref}

\usepackage[dvipsnames]{xcolor}
\usepackage{amsmath}
\usepackage{graphicx}
\usepackage[title]{appendix}
\usepackage{mathrsfs}
\usepackage{authblk}
\usepackage{amsfonts}
\usepackage{booktabs} 
\usepackage{caption}  
\usepackage{threeparttable} 
\usepackage{algorithm}
\usepackage{algorithmicx}
\usepackage{algpseudocode}
\usepackage{listings}
\usepackage{enumitem}
\usepackage{chngcntr}
\usepackage{booktabs}
\usepackage{lipsum}
\usepackage{subcaption}
\usepackage{authblk}
\usepackage[T1]{fontenc}    
\usepackage{csquotes}       
\usepackage{diagbox}
\usepackage{babel}
\usepackage{multirow}
\usepackage{array}
\usepackage{caption}
\usepackage{graphicx}
\usepackage[utf8]{inputenc}
\usepackage{newunicodechar}
\newunicodechar{−}{\textminus}
\usepackage[british]{babel}
\usepackage{setspace}
\usepackage{fancyhdr}
\newcommand{\bx}{\boldsymbol{x}}
\newcommand{\bp}{\boldsymbol{p}}

\newcommand{\balpha}{\boldsymbol{\alpha}}

\newcommand{\btheta}{\boldsymbol{\theta}}
\newcommand{\heightdrift}{\text{height}_\text{drift}}
\newcommand{\modedrift}{\text{mode}_\text{drift}}
\newcommand{\IQRdrift}{\text{IQR}_\text{drift}}

\begin{document}

\begin{titlepage}
    \centering
    \doublespacing
    \vspace*{\fill}
    {\LARGE Extending evidence accumulation models to bounded continuous self-report data} \\[1.5cm]
    {\large \text{Yufei Wu}$^{1,*}$, \text{Tamás Szűcs}$^1$, 
    \text{Agnes Moors}$^1$, \text{Francis Tuerlinckx}$^1$} \\[0.5cm]
    {\small $^1$KU Leuven, Belgium} \\[0.2cm]
    {\small Corresponding author: \texttt{yufei.wu@kuleuven.be}}
    \vspace*{\fill}
\end{titlepage}

\begin{abstract}
\noindent Evidence accumulation models (EAMs) provide a powerful framework for inferring latent cognitive processes from choice and reaction time data. While EAMs are traditionally limited to binary choices, recent developments have extended them to rotationally symmetric continuous responses via the circular diffusion model \citep{smith2016diffusion} and the spatially continuous diffusion model \citep{ratcliff2018decision}. Yet, such extensions are limited in scope, as many psychological constructs are measured on bounded non-rotational scales. In this paper, we bridge this gap by presenting and comparing two adaptations designed for bounded continuous data: the Half-Circular Diffusion Model (HCDM) and the Beta Drift Diffusion Model (BDDM). Because both models have intractable likelihoods, we fit them using Amortized Bayesian Inference (ABI) and compare them using Amortized Bayesian Model Comparison (ABMC). We demonstrate the complete workflow on an empirical affect dataset (N = 215), including parameter recovery, simulation-based calibration, posterior predictive checks, and model comparison. Both models accurately capture the joint distribution of responses and reaction times and yield interpretable parameters that can be reliably recovered. The model comparison further reveals a simple diagnostic for choosing between them: the dispersion of the rating distribution, with HCDM preferred for moderate spread and BDDM for highly concentrated or highly dispersed ratings. This work extends the EAM framework to a new application context, bounded continuous self-report data, and offers researchers a user-friendly toolkit for modeling the cognitive dynamics of continuous responses. We release fully documented Python code with both GPU and CPU implementations, along with example datasets.
\end{abstract}

\textbf{Keywords}: evidence accumulation models, self-report data, bounded continuous data, amortized Bayesian inference, amortized Bayesian model comparison

\section*{Introduction}

\begin{figure}[!ht] 
 \centering
 \includegraphics[width=1\linewidth]{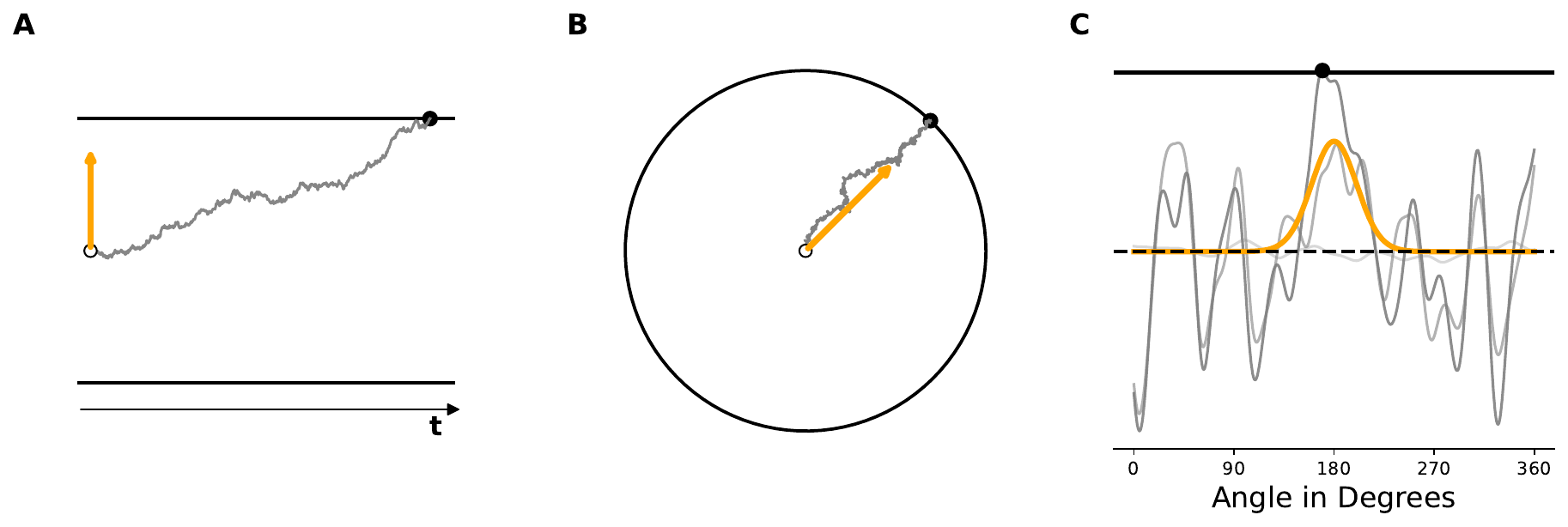}
 \caption{ Schematic illustration of the DDM, the CDM, and the SCDM. In all panels, the orange line indicates the average rate of evidence accumulation. In the DDM (A), the drift rate is one-dimensional, and the grey line shows evidence accumulating over time until it reaches one of the two horizontal boundaries, triggering a decision. In the CDM (B), both drift rate and accumulation are two-dimensional, so time is not directly shown but implicit; a decision occurs when evidence reaches the circular boundary. In the SCDM (C), the drift rate follows a Gaussian profile across the decision space, and evidence accumulates over the entire space. Snapshots at different time steps (grey curves) illustrate this process: Darker curves indicate more recent evidence, and the darkest curve shows the moment the boundary is reached. Evidence can be negative (below the dashed line) because it is normalized to zero at each time step.} 
 \label{fig:oldmodels}
\end{figure}

Evidence accumulation models (EAMs) have become one of the most important modeling frameworks for speeded decision-making in the past decades \citep{ratcliff1978theory, evans2020evidence}. The core assumption of EAMs is that evidence in favor of each decision option accumulates over time until the evidence for one option reaches a decision boundary, at which point a decision is triggered. This framework enables researchers to jointly model choices and their associated reaction times. For example, the drift diffusion model \citep[DDM;][]{ratcliff1978theory}, one of the most prominent instances of EAMs, assumes that noisy evidence accumulates at an average rate $v$, following the stochastic process $dX_t = v \, dt + \sigma \, dW_t$, where $dW_t$ denotes a Wiener process representing random noise. A decision is made at time $t$ when the accumulated evidence $X_t$ crosses one of the two decision boundaries (e.g., $X_t \ge a$ or $X_t \le 0$). Each boundary represents a specific decision (Figure~\ref{fig:oldmodels}, panel A). Over the years, the general EAM framework has given rise to numerous model variants, either suggesting novel theoretical accounts of decision-making \citep{brown2008integrated} or offering simplified measurement tools for applied research \citep{brown2008simplest}.

As theoretical accounts of decision-making, EAMs formalize how evidence is accumulated over time, often achieving a close quantitative fit to empirical choice reaction time distributions across a wide range of paradigms \citep{evans2020evidence}. EAMs decompose the choice reaction time distribution into psychologically meaningful latent variables. In an experiment where EAMs are applied, participants are asked to make binary decisions, such as determining the random dots movement direction (left or right) \citep[]{mulder_bias_2012}, classifying words and non-words \citep[]{ratcliff_diffusion_2004}, or reporting affect (positive or negative) \citep{szucs_wu_tuerlinckx_moors2026}. Depending on the exact paradigm, the drift rate can reflect the salience of stimuli \citep{ratcliff_modeling_1998} or the ability of participants \citep{von_krause_mental_2022}; the decision boundary representing response caution, can index the trade-off between speed and accuracy \citep{ratcliff_modeling_1998, von_krause_mental_2022}. Performance is then compared across different conditions (e.g., primed vs. non-primed) or participant demographics (e.g., younger vs. older adults).

This work aims to extend EAMs to bounded, continuous self-report data. A common way to measure constructs such as emotion intensity \citep{mauss2009measures}, risk preference \citep{charness2013experimental}, and confidence \citep{kleitman2007self} is through visual analogue \citep{gift1989visual} or Likert-type scales with many response levels \citep{clark1995constructing}. Importantly, such scales are intended to approximate continuous measurement and differ fundamentally from coarse categorical responses (e.g., three or four response levels), which may require distinct modeling approaches \citep{trueblood2014multiattribute}. A second defining property of these data is that responses are bounded with meaningful lower and upper endpoints (e.g., 1 to 100 on the scale could represent ``not at all'' to ``very much''). The third key property is that the data are subjective self-reports, which means that there is no externally specified, correct location along the continuum to which the evidence accumulates. Modeling such approximately continuous, bounded self-report responses jointly with their reaction times poses challenges for conventional EAMs but also opens new opportunities to study a wide range of psychological processes within a unified decision-theoretic framework.

Two EAMs have been proposed and implemented for continuous response tasks: the circular diffusion model \citep[CDM;][]{smith2016diffusion} and the spatially continuous diffusion model \citep[SCDM;][]{ratcliff2018decision, ratcliff2024using}. A third framework based on a deterministic, rather than stochastic, evidence accumulation process has also been proposed \citep{kvam2023unified}; however, in the present work, we focus on stochastic evidence accumulation processes. The CDM extends one-dimensional diffusion in DDM to a two-dimensional circular decision space, where evidence begins at the origin and evolves stochastically until it reaches a circular boundary (see panel B in Figure~\ref{fig:oldmodels}). This makes the CDM suitable for tasks with rotationally symmetric response domains but less appropriate for bounded data. The SCDM models evidence accumulation over a continuous spatial domain (i.e., a circle or a line), assuming that the drift rate follows a Gaussian-shaped profile across the decision space (Figure~\ref{fig:oldmodels}, panel C). This profile is defined by its height (reflecting evidence strength) and width (reflecting uncertainty), and its center corresponds to the spatial location of the fastest evidence accumulation—known in each trial based on the stimulus presented. This makes the SCDM well-suited for perceptual estimation tasks where the true stimulus value is externally specified. However, in subjective self-report paradigms, the correct location of evidence is not externally given and must instead be inferred from the responses themselves, a scenario not supported by the current SCDM formulation. 

In the present work, we adapt both the CDM and the SCDM to accommodate bounded continuous self-report data. The resulting models lack tractable likelihoods, rendering conventional parameter estimation and model comparison approaches infeasible. To address this challenge, we leverage recent advances in likelihood-free model fitting and comparison, specifically amortized Bayesian inference \citep[ABI;][]{radev_bayesflow_2020} and amortized Bayesian model comparison \citep[ABMC,][]{radev2021amortized, elsemuller2024deep}. ABI is a likelihood-free approach that trains neural networks to learn the mapping between data and their corresponding posterior distributions, using large sets of simulated data paired with known ground-truth parameters. ABMC works by training neural networks with simulated data with known model labels, framing model comparison as a probabilistic classification task. To demonstrate this simulation-based workflow (i.e., simulation, parameter recovery, estimation, posterior predictive checking, and model comparison), we apply both models to an empirical dataset featuring bounded continuous self-report responses. 

Our contribution is threefold. First, we formalize and implement two stochastic EAMs, the HCDM and the BDDM,  for bounded continuous self-report data, completing a gap in the EAM family that has so far been limited to binary or rotationally symmetric response domains. Second, we demonstrate that recent advances in ABI and ABMC, making these otherwise intractable models tractable for routine data analysis, with parameter recovery, simulation-based calibration, posterior predictive checks, and model comparison all completing in minutes on a single GPU. Third, we deliver a complete, reproducible workflow that researchers can adopt for their own bounded continuous response data. Together, these contributions transform a class of models that has been theoretically motivated but practically inaccessible into a usable analytic toolkit for experimental psychology.

\section*{The half-circular diffusion model (HCDM)}

\paragraph{The original circular diffusion model.} Before moving to the half-circular version, let us start with the circular one, the CDM. The CDM assumes evidence accumulation within a bounded disc centered at the origin and with radius $a$. Let $\boldsymbol{x}_t=(x_{1t},x_{2t})^\top$ denote the current evidence state, which is a two-component vector. The model for the dynamics of the evidence is a two-dimensional stochastic process: 
\begin{equation*}
    d \boldsymbol{x}_t = \boldsymbol{v}\, dt + \boldsymbol{\sigma} d\boldsymbol{w}_t ,
\end{equation*}
where $\boldsymbol{w}_t$ denotes the standard Brownian motion, and it is scaled by its standard deviation $\boldsymbol{\sigma} = \sigma I_2$ (with $I_2$ being the $2\times 2$ identity matrix). For convenience and ease of comparison with the half-circular version later, we will also present the discrete-time approximation of the model:
\begin{equation}
    \boldsymbol{x}_{t+\Delta t} = \boldsymbol{x}_t + \boldsymbol{v}\, \Delta t + \boldsymbol{\sigma}\boldsymbol{w}_t \sqrt{\Delta t} ,
    \label{eq:hcdmsim}
\end{equation}
with $\Delta t$ being a small time step, $\boldsymbol{w}_t$ a bivariate normally distributed random vector, and $\boldsymbol{\sigma}$ set to be $ \left(\begin{smallmatrix} 1 & 0 \\ 0 & 1 \end{smallmatrix}\right)$.

The relative starting point is $\boldsymbol{z} = (z_1, z_2)^\top$, where $z_1, z_2 \in [-1, 1]$. The evidence accumulation process starts from $(z_1 \, a, z_2 \, a)$ and evolves within the disc centered at the origin and with radius $a$: $\mathcal{D} = \{(x_1,x_2): x_1^2 + x_2^2 \leq a^2\}$. In Smith (\citeyear{smith2016diffusion}), the starting point for evidence accumulation is constrained to $(0,0)$, which together with the rotational symmetry of the disc yields a tractable likelihood.

A decision is registered when the accumulated evidence first crosses the arc: $x_{1t}^2 + x_{2t}^2 = a^2$. The first-passage time defines the time of the evidence accumulation process: $T = \inf \{ t \geq 0: x_{1t}^2 + x_{2t}^2 = a^2 \}$. The observable response or reaction time is obtained by adding to $T$ the non-decision time $T_{er}$ (e.g., time for stimulus encoding and motor execution): $RT = T+T_{er}$. The polar angle at the position of the boundary crossing is defined as $\phi=\mbox{atan2}\left(\frac{x_{2t}}{x_{1t}}\right)$\footnote{The $\mbox{atan2}$ is defined in such a way that it returns a reasonable result for any possible combination of $x_{1t}$ and $x_{2t}$ (also e.g., when $x_{2t}=0$).}. This angle $\phi$ is measured from the origin in a counterclockwise direction to the first-passage point, and it is assumed to be the response $Y=\phi$. Thus, $Y \in (-\pi,\pi]$.

\paragraph{The half-circular diffusion model.} To adjust this model so that it may be fitted to bounded continuous self-report data, we need to confine the process to $\mathcal{D}^{\prime} = \{(x_1,x_2): x_1^2 + x_2^2 < a^2, \, x_2 \ge 0\}$, a bounded semi-disc, resulting in the HCDM. The starting point $\boldsymbol{z} = (z_1, z_2)^\top$ now has components: $z_1 \in [-1, 1]$ and $z_2 \in [0, 1]$.  In order to confine the evidence accumulation process to the half circle, a specular reflection is applied at the diameter $x_2=0$, preserving the geometric direction of motion while preventing the trajectory from leaving the valid region (see the right panel in Figure~\ref{fig:HCDM}, and Equation~\ref{eq:reflection} in the Appendix). The conceptual basis of the HCDM, including the reflecting boundary mechanism, was first described by Smith (\citeyear{smith2016diffusion}) as a potential extension of the CDM. In the present work, we formalize this idea, elaborate on its implementation (specifically for self-report data), and provide the first empirical application.

\begin{figure}[!ht] 
 \centering
 \includegraphics[width=1\linewidth]{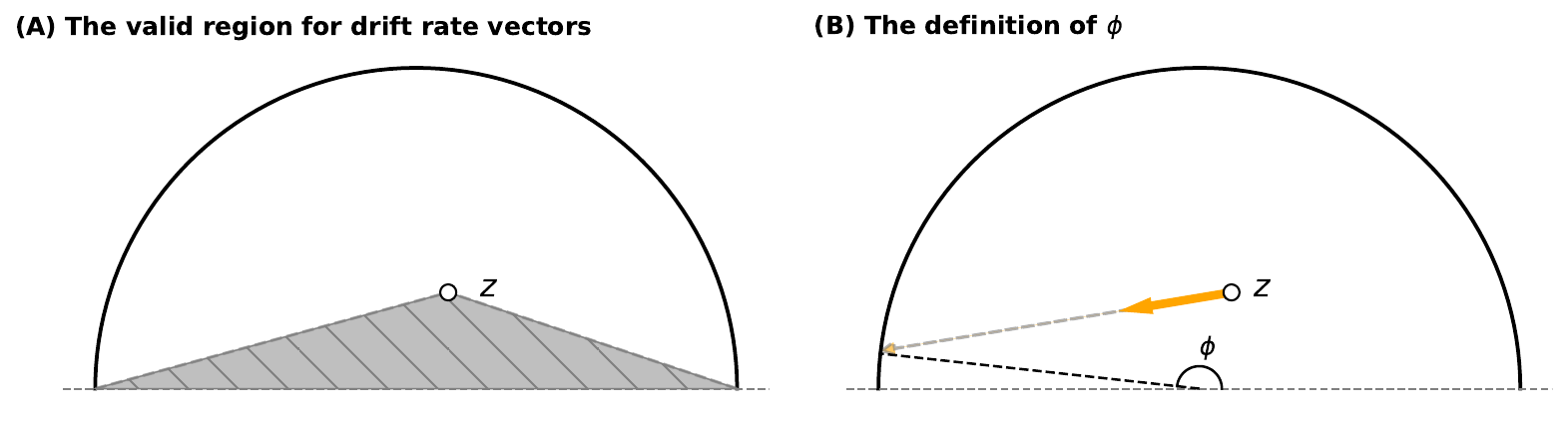}
 \caption{Schematic illustration of the valid region for drift rate vectors and the definition of $\phi$ in HCDM. Panel A shows that, when $z$ moves away from the origin, the valid region (the white region) into which the drift rate vectors can point changes accordingly. Panel B shows how $\phi$ is defined.} 
 \label{fig:HCDMprior}
\end{figure}

Importantly, although the HCDM is defined using a vector-component parameterization for notational convenience, the drift rate vector can be parameterized in polar coordinates. One challenge arises when $z$ moves away from the origin: the valid region (white region) into which the drift rate vector can point changes accordingly (Figure~\ref{fig:HCDMprior}, left panel). As a result, we parameterize drift rate vectors as $(l, \phi)^\top$, where $l$ denotes the magnitude of the drift rate, and the angle $\phi$ is defined by the intersection of the decision arc with the ray extending from the origin in the direction of the drift rate (see the right panel in Figure~\ref{fig:HCDMprior}). The definition ties $\phi$ to positions along the arc rather than to a fixed coordinate axis. As a result, the valid range of $\phi$ is independent of the relative starting point $z$, allowing $z$ to vary away from $(0,0)^\top$ \footnote{While the drift rate vector can be represented in either Cartesian or polar coordinates, the choice is a matter of utility. In the provided scripts, we utilize Cartesian coordinates for computational efficiency during simulation, whereas polar coordinates are used in psychological interpretation.}. Crucially, this parameterization also allows responses on the bounded scale to be directly mapped on $\phi \in [0, \pi]$ through a linear rescaling.
\begin{figure}[!ht] 
 \centering
 \includegraphics[width=1\linewidth]{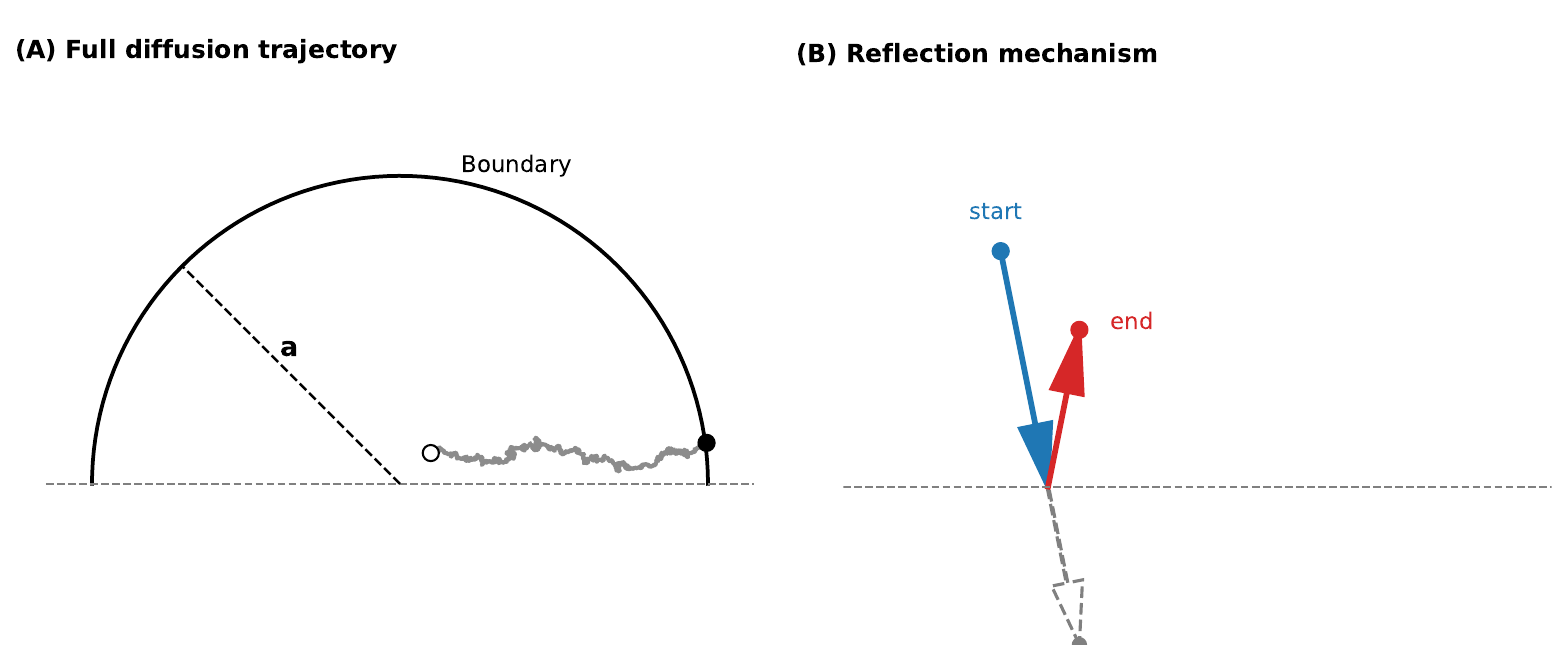}
 \caption{Schematic illustration of the HCDM. Panel A illustrates a complete diffusion trajectory, where the two-dimensional diffusion process starts from the relative starting point $(0.1, 0.1)$ and terminates at a point on the semicircular boundary of radius $a = 1.0$.
Panel B zooms in on the reflection mechanism that occurs when the diffusion process is about to cross the flat edge at $x_2 = 0$. The blue segment represents the portion of the step that reaches the boundary, and the grey dashed segment indicates the remaining part of the step. This remaining displacement is reflected with respect to the boundary, producing the red segment that continues the trajectory within the valid region.} 
 \label{fig:HCDM}
\end{figure}

\section*{The beta drift diffusion model (BDDM)}

\paragraph{The original spatially continuous diffusion model.} The SCDM assumes that evidence accumulates over a continuous spatial domain. In the SCDM, the drift rate is a (scaled, possibly non-normalized) Gaussian-shaped function \footnote{Since the scale is rotationally symmetric, the drift rate is realized by a scaled von Mises distribution, the circular analogue of the normal distribution. In SCDM, the drift rate is always centered at 180 degrees, representing the correct response (see panel C in Figure~\ref{fig:oldmodels}, where the x-axis is angle in degrees). Ratcliff (\citeyear{ratcliff2018decision}) notes that with this representation, the Gaussian and von Mises distributions are indistinguishable.}. Therefore, it can be considered a (continuous) drift rate function, because it is defined simultaneously for all locations across the decision space. It has three parameters that can be inferred from the data: the height, the standard deviation, and the trial-to-trial variability in height (a uniform range). It is important to note that the center (i.e., mean) of the drift rate function is not a free parameter inferred from the data, but is instead determined by the external stimulus. This formulation was developed for perceptual decision-making tasks in which the true stimulus value is known. For example, in tasks where participants identify the dominant color in a central patch by moving their eyes to the corresponding location on a surrounding circular annulus \citep{ratcliff2018decision}, the center of the drift rate function corresponds to the location of the dominant color on the annulus. The stochastic component in this evidence accumulation process is modeled as the addition of the drift rate function perturbed at each time instant by a spatial Gaussian process on the same spatial domain \citep{ratcliff2018decision}. The equation for the evidence accumulation process then becomes: 
\begin{equation}
    x_{t+\Delta t} (u) = x_t(u) + v(u)  \Delta t + \sigma  \eta_t(u)  \sqrt{\Delta t}, \label{eq:bddm}
\end{equation}
where $u$ is the spatial variable (e.g., $u \in [0,2\pi]$, $t$ is the time, $\Delta t$ is the time step, and $\eta_t(u)$ is the Gaussian process noise at time $t$. When the accumulated evidence exceeds the boundary ($x_t(u) \geq a$), the response and reaction time are registered as $u$ and $T + T_{er}$.

Importantly, using a Gaussian process to model noise $\eta(u)$ ensures that the noise at neighboring locations is correlated, preventing spatially independent noise to produce jittery, biologically implausible accumulation. Consider $u$ and $u'$ as two points in an interval $[0, 1]$. Then covariance $K$ between these points is defined by a squared exponential kernel: 
\begin{equation}
    K(u,u') = \exp\left(-\frac{(u-u')^2}{2\rho^2}\right).
    \label{eq:kernel}
\end{equation}
The parameter $\rho$ controls the smoothness of the noise: the larger the $\rho$, the more the points $u$ and $u'$ are correlated (and vice versa). 

\begin{figure}[!ht] 
 \centering
 \includegraphics[width=1\linewidth]{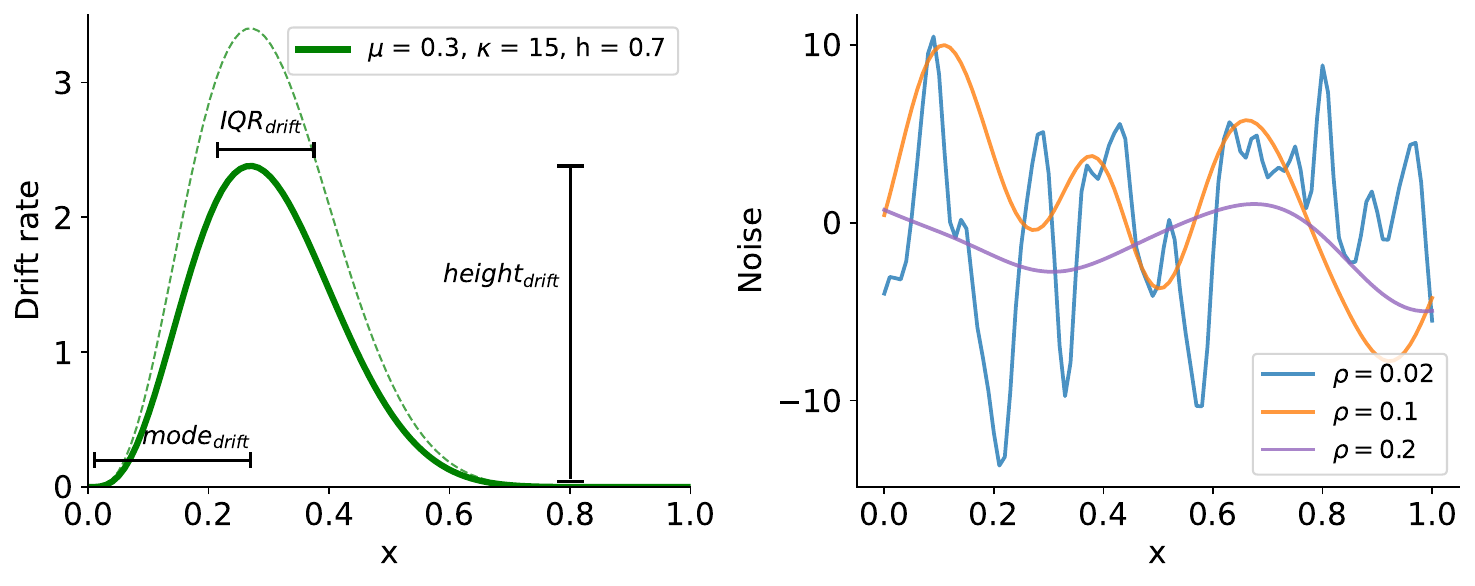}
 \caption{Schematic illustration of the drift rate and noise processes in BDDM. The left panel shows a drift rate profile. The dashed line represents the PDF of a beta distribution. The bold green line is the drift rate function, a beta density scaled by $h$. The ($\heightdrift$, $\modedrift$, $\IQRdrift$) are the geometric features of the drift rate function. The right panel shows three random Gaussian process noise functions with different kernel parameters $\rho$ that control their smoothness.} 
 \label{fig:BDDM}
\end{figure}

To simulate such a Gaussian process $\eta(u)$ for practical data analysis, we discretize the spatial domain into a regular grid: $u_i = i \Delta u$ for $i=0,\dots,n$, where $\Delta u$ is chosen such that $n\Delta u=1$. For example, in the right panel of Figure~\ref{fig:BDDM}, we took $n = 100$ and $\Delta u = 0.01$. The covariance matrix $\Sigma_{\Delta u}$ thus has entries $\sigma_{ij} = K(u_i,u_j)$. For efficient computation, $\Sigma_{\Delta u}$ is decomposed to a lower triangle matrix and its conjugate transpose $\Sigma_{\Delta u}=LL^\top$. By drawing a vector of $n$ independent samples $\boldsymbol{\epsilon} \sim \mathcal{N}(0,1)$ and projecting these samples into the correlated spatial domain as follows: 
\begin{equation*}
    \boldsymbol{\eta} = \boldsymbol{z} L^\top,
\end{equation*}
a Gaussian process is realized, where the vector $\boldsymbol{\eta} = (\eta(u_0),\dots,\eta(u_n))^\top$, preserving the spatial dependency defined by the kernel. For more information, see \citep{ratcliff2018decision, lord2014introduction}. 

 An important assumption behind EAMs is that at each time point, the evidence for one response is evidence against the other. Thus, the total amount of accumulated evidence is constant across time (normalized to zero) \citep{ratcliff2018decision}.

\paragraph{The beta drift diffusion model.} We modify the parameterization of the drift rate function to a scaled beta probability density function (PDF) to accommodate for the bounded nature of most continuous self-report data. The PDF of a beta distribution is parameterized by a mean $\mu$ and a concentration $\kappa$: $f(u) = \frac{u^{\mu \kappa - 1}(1-u)^{\kappa - \mu  \kappa-1}}{\text{B}(\mu  \kappa, \kappa - \mu  \kappa)}$, where $\text{B}(\mu  \kappa, \kappa - \mu  \kappa) = \frac{\Gamma(\mu  \kappa)\Gamma(\kappa - \mu  \kappa)}{\Gamma(\mu  \kappa + \kappa - \mu  \kappa)}$ and $\Gamma(\cdot)$ is the Gamma function. To release the constraint that the integral of a PDF is always 1, we scale the PDF by $h$, so that the drift rate function is $h \cdot f(u)$ for $u \in [0, 1]$ (see the left panel of Figure~\ref{fig:BDDM}). Note that while the model is defined on [0,1], continuous self-report data with finite lower and upper bounds can be rescaled linearly to this interval (or vice versa). To simulate from the model and perform calculations, we consider the grid size $n$ (i.e., determining the degree of approximation) to be 100. This means that there are $n+1=101$ accumulators considered in parallel. 

Although the drift rate function is formally parameterized by $(\mu, \kappa, h)$, we also characterize it via its key geometric features: the mode  $(\text{mode}_\text{drift})$, the inter-quartile range $(\text{IQR}_\text{drift})$, and the peak height $(\text{height}_\text{drift})$ to facilitate interpretation. The mode was chosen because it directly shows the peak location of the geometric shape (which is the most likely region that will cross the threshold first). The interquartile range of the drift rate function, $\text{IQR}_\text{drift}$, is a measure of how concentrated it is \footnote{$IQR = Q_3 - Q_1$, where $Q_1$ is the first quartile (i.e., 0.25 quantile) and $Q_3$ is the third quartile (i.e., 0.75 quantile).}. A smaller IQR means the central part of the curve is more concentrated. IQR is more informative than the concentration parameter $\kappa$ when the drift rate function is highly skewed. Lastly, the maximum height of the drift rate function (fastest evidence accumulation speed across the decision space), denoted as $\text{height}_\text{drift}$, provides a direct and intuitive measure of the magnitude. By mapping the abstract parameter set to these geometric properties, we preserve the interpretation of the drift rate as a continuous average rate of accumulation, while providing more transparent links to observed behavior.

\begin{table}[h]
\centering
\begin{tabular}{cllc}
\toprule
Model & \multicolumn{2}{l}{Parameter} & Notation  \\
\midrule
    HCDM    & Drift rate vector       & length & l \\
            &                         & angle  & $\phi$ \\
            & Boundary                &        & a \\
            & Starting point          &        & $(z_x, z_y)$ \\
            & Non-decision time       &        & $T_{er}$\\
\midrule
    BDDM    & Drift rate function     & height        & $\text{height}_\text{drift}$ \\
            &                         & mode          & $\text{mode}_\text{drift}$ \\
            &                         & concentration & $\text{IQR}_\text{drift}$ \\
            & Boundary                &               & a  \\
            & Kernel parameter        &               & $\rho$ \\
            & Non-decision time       &               & $T_{er}$\\
\bottomrule
\end{tabular}
\caption{The parameters or geometric features in HCDM and BDDM.}
\label{tab:partab}
\end{table}

\section*{Application}

We used an empirical dataset to illustrate how the two models can be applied to bounded continuous self-report data with response times. In recent years, researchers have proposed framing self-reported affect as a decision-making process \citep{teoh_framing_2023, givon_are_2023} and have applied EAMs on binary choice reaction time data for theoretical development and hypothesis testing. However, the most commonly used response scales in affect research are not binary, but Likert or continuous \citep{ekkekakis2013measurement}.  Therefore, in line with the earlier EAM research on affect, we show how to analyze a set of bounded continuous self-report affective data under the EAM framework.  Code for conducting all the analyses and reproducing all results is available at \url{https://github.com/yufeiwu1011/EAMs-for-Bounded-Continuous-Data}.

\subsection*{Experiment}

\begin{figure}
\centering
\includegraphics[width=0.9\linewidth]{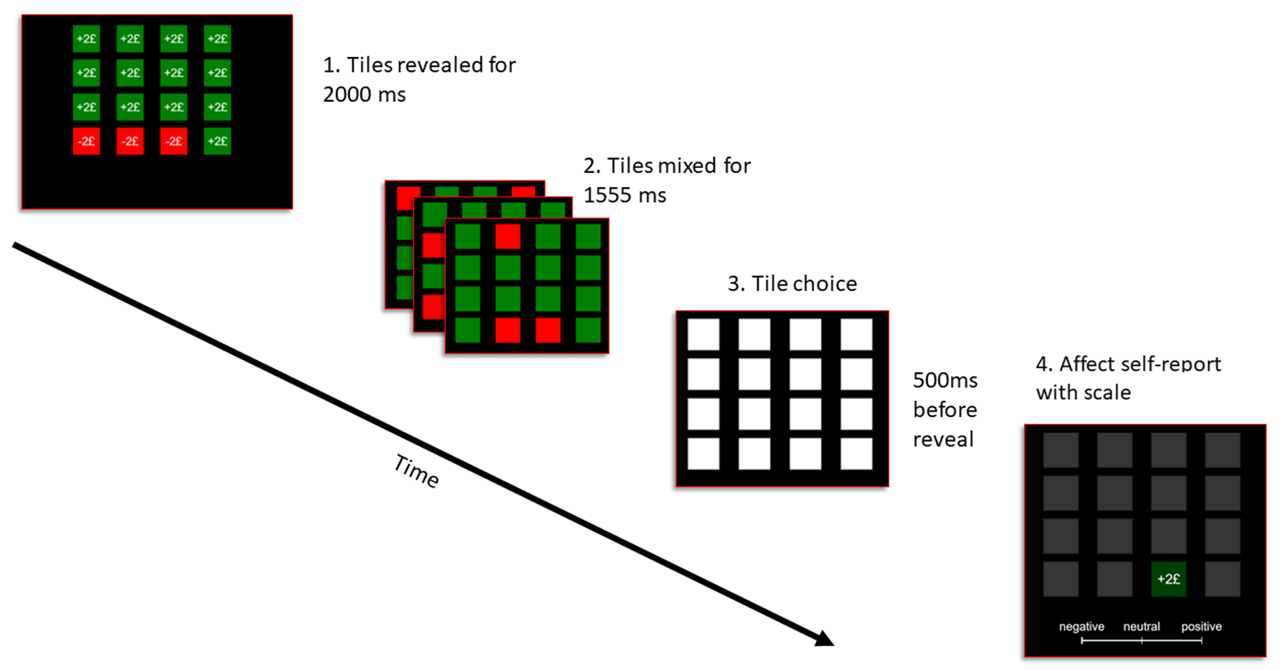}
\caption{\label{fig:trial}Participants chose from tiles hiding monetary outcomes (e.g., 3 with +£2, 13 with –£1). After revealing the outcome, they reported their subjective affect on a continuous scale that appeared at the bottom.}
\end{figure}

The data were collected for a different study using a probabilistic reward task, originally designed to investigate the determinants of affect \citep{szucs_wu_tuerlinckx_moors2026}. Figure~\ref{fig:trial} illustrates the flow of a single trial. Participants were presented a $4 \times 4$ grid of possible outcomes (e.g., −£2, +£1) for 2000 ms. After the grid was shuffled and the outcomes hidden, participants selected one square to reveal their actual win or loss. Following the outcome, participants reported their affective state using a visually continuous slider (i.e., visual analogue scale) anchored by ``negative'', ``neutral'', and ``positive''. The scale was implemented as a line, with its length in pixels adjusted to the resolution of the participant’s screen. For example, on a 1920-pixel-wide display, the scale length was 192 pixels, allowing for high-resolution measurement of the response. Both the final rating (normalized to  $[0, \pi]$ for the analysis) and the response time (defined as the latency from the scale's appearance to the final click) were recorded. 

This pseudo-gambling task used a $4 \times 6$ within-subject design with two factors: outcome (four levels: −£2, −£1, +£1, +£2) and win probability (six levels: $\frac{1}{16}$, $\frac{2}{16}$, $\frac{3}{16}$, $\frac{13}{16}$, $\frac{14}{16}$, $\frac{15}{16}$). Each condition was repeated 8 times, totaling 192 trials per participant, plus 4 attention check trials. Participants were explicitly instructed that trials were independent and were asked to respond as quickly and accurately as possible. For full details of the paradigm, see Szűcs et al. (\citeyear{szucs_wu_tuerlinckx_moors2026}). 

Note that although the experiment also manipulated win probability, we ignored this variable in model fitting because (a) including it would overly complicate the model given the limited number of trials per participant, and (b) linear model analyses of the dataset showed that outcome explained more variation in ratings than win probability \citep{szucs_wu_tuerlinckx_moors2026}. After aggregating trials with the same outcome but different probabilities into a single condition, each condition contained 48 trials.

We excluded trials with reaction times below 100 ms or above 5000 ms, as well as participants with more than 10\% of trials excluded or who failed more than two attention checks. After applying these criteria, 215 of the 250 recruited participants were retained for further analyses.

\subsection*{Model fitting}

\subsubsection*{Method}

Since the likelihoods of the HCDM and BDDM are intractable, we opt for ABI, a likelihood-free inference method. ABI utilizes model simulations to train specialized neural networks that learn to compress data of varying sizes and to sample from the posterior over model parameters given \textit{any} data set compatible with the model \citep{zammit2024neural}. Two networks are involved: (a) A summary network that transforms each data set into fixed-size \textit{approximately sufficient} summary statistics. (b) Simultaneously, the inference network learns to approximate the true posterior of model parameters given the data. Both models are fitted using the ABI method implemented in the \texttt{BayesFlow} package in Python \citep{radev_bayesflow_2020}. For a detailed ABI workflow, see the Appendix.

\subsubsection*{HCDM}

When fitting the HCDM, we assumed that the four monetary outcomes (−£2, −£1, +£1, +£2) correspond to four drift rates, while the other parameters remained constant across conditions. 

The first step in model fitting is simulating training data from HCDM. Ground-truth parameters were sampled from the following priors: $l \sim U(0.5, 9)$, $\phi \sim U(0, \pi)$, $z_1 \sim U(-0.7, 0.7)$, $z_2 \sim U(0, 0.7)$, $a \sim U(0.5, 9)$, and $T_{er} \sim U(0.1, 2.5)$. We then simulated decision-making datasets based on the priors. The simulations were performed using the Euler–Maruyama method, with a $\Delta t$ of $0.001\,s$ (see Equation~\ref{eq:hcdmsim}). The trial number per condition was sampled uniformly from the integers 39 through 50, matching the varying trial numbers in the empirical data.

\begin{figure}[!ht] 
 \centering
 \includegraphics[width=1\linewidth]{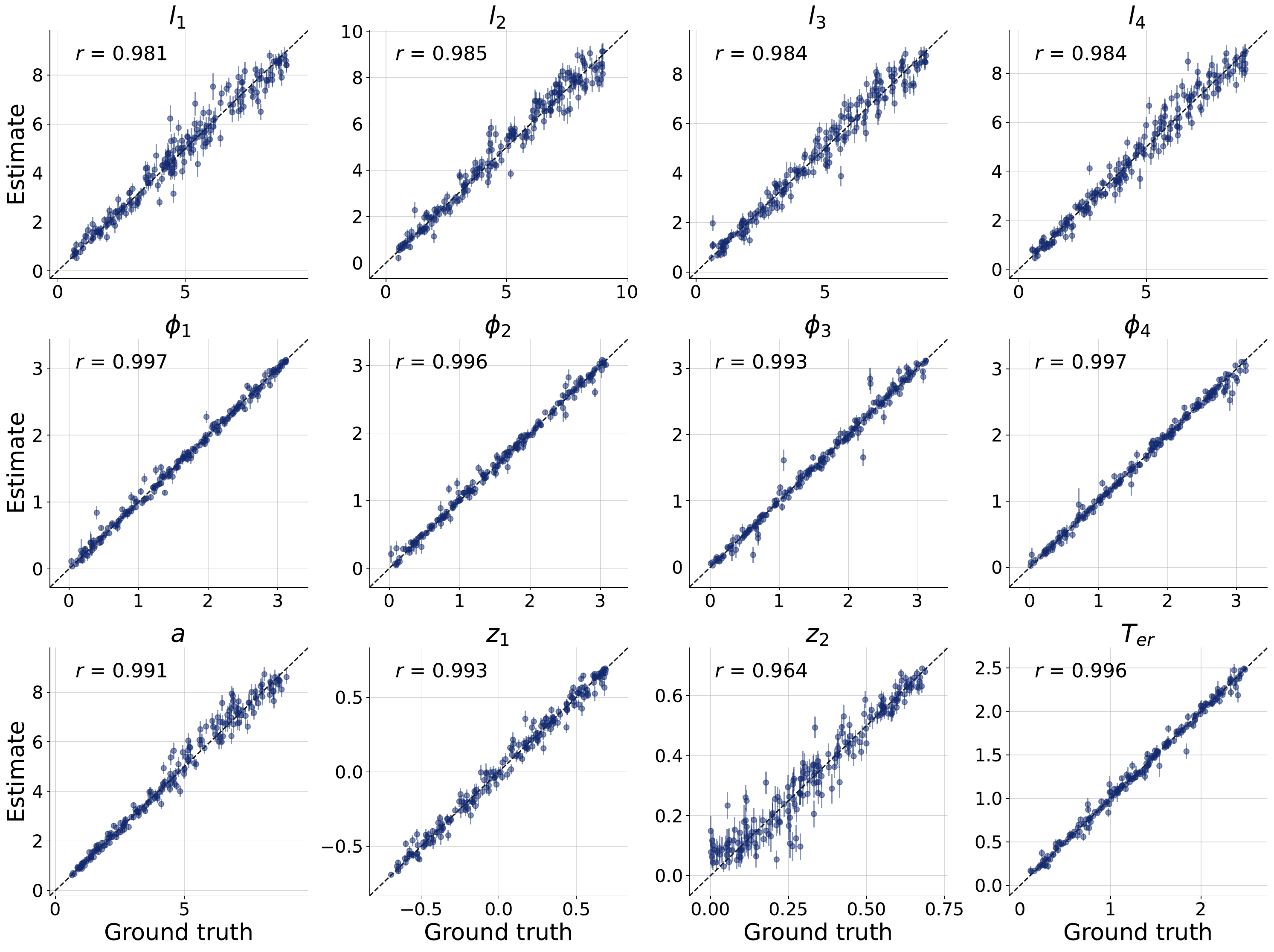}
 \caption{The parameter recovery of the HCDM. The x-axis stands for the ground truth parameter values, and the y-axis represents their estimates. The black dashed line is the identity line (indicating perfect recovery). In the top left corner, the correlation between true and estimated values is shown.} 
 \label{fig:HCDMrecovery}
\end{figure}

For the neural network architecture, we used a SetTransformer \citep{lee2019set} with a 128-dimensional output as the summary network, and a conditional flow matching network with a 4-layer MLP (256 units per layer) as the inference network \citep{lipmanflow}. Training was performed on an NVIDIA Tesla H100-SXM2-32GB GPU. The model was trained for 100 epochs (512 batches per epoch, batch size 64, online training) using the Adam optimizer (initial learning rate $5\times 10^{-4}$, with a cosine decay schedule), and the loss had converged by the end of training. Each epoch took approximately 6 seconds, resulting in a total training time of 10 minutes. We further simulated 200 additional validation datasets to assess parameter recovery and found that all parameters were well-recovered as can be seen in Figure~\ref{fig:HCDMrecovery}. In addition,  the simulation-based calibration shows that the estimates were unbiased (see Figure~\ref{fig:ecdf_hcdm}).

\begin{figure}[!ht] 
 \centering
 \includegraphics[width=1\linewidth]{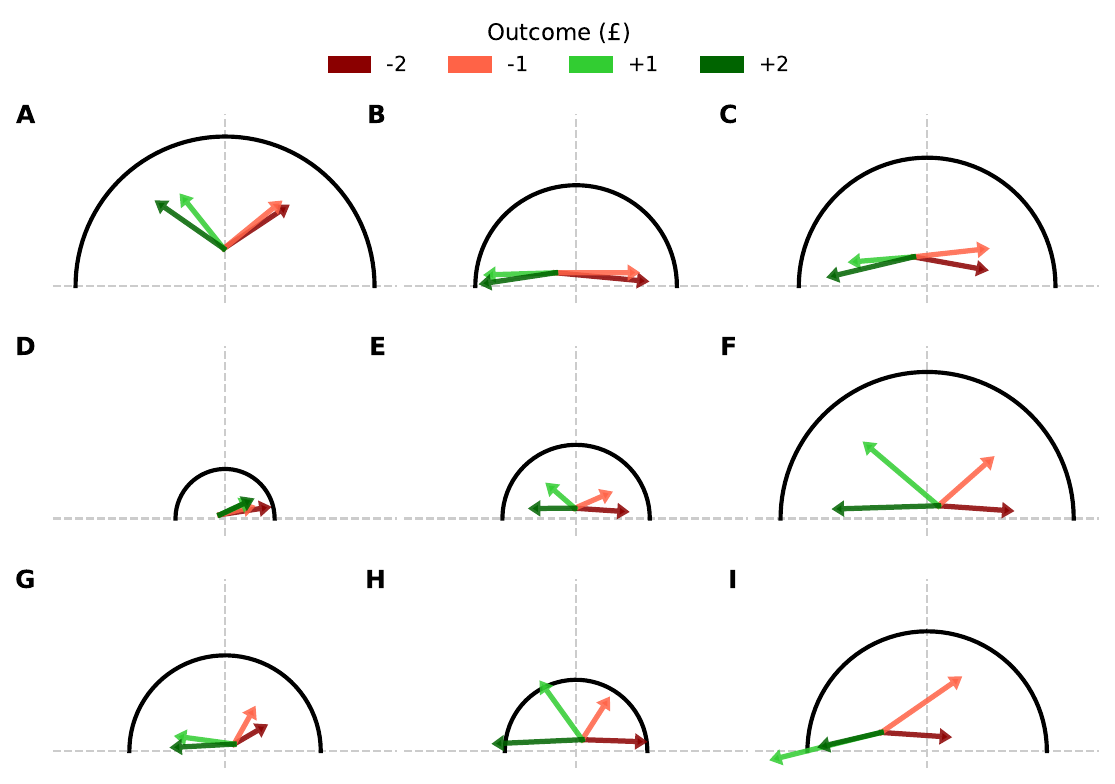}
 \caption{Visualization of the posterior means of drift rate vectors and boundary for nine participants in the HCDM. The black arc represents the decision boundary $a$, the colored arrows stand for the drift rate $(l, \phi)$, and the starting point of the arrows is the starting point of the evidence accumulation process.} 
 \label{fig:HCDMfitting}
\end{figure}

We then fitted the empirical data to the model by inputting each participant's data into the trained neural networks to obtain posterior samples. Posterior sampling for all 215 participants took approximately 376 milliseconds (1000 samples per participant). Figure~\ref{fig:HCDMfitting} visualizes the posterior means of drift rate vectors and boundaries for nine randomly selected participants. While the radius of the half-circle represents the decision boundary, the colored arrows indicate the four drift rate vectors corresponding to the four outcome conditions \footnote{In some subplots (e.g., participants H, and I), the drift rate vectors appear to cross the boundaries. This does not imply that the evidence accumulated in a single time step; rather, the average evidence accumulation magnitude per time step is $l \times \Delta t$.}. 

Except for participant D, participants generally showed a clear relationship between outcomes and decisions: Lower outcomes were associated with smaller $\phi$, indicating lower affect ratings. The evidence accumulation speed, parameterized by the drift rate length $l$, varied across conditions without a consistent pattern. For example, for participants F, G, and H, $l$ was larger in the winning conditions than in the losing conditions, whereas this pattern was not observed in the other participants.

\begin{figure}[!ht] 
 \centering
 \includegraphics[width=1\linewidth]{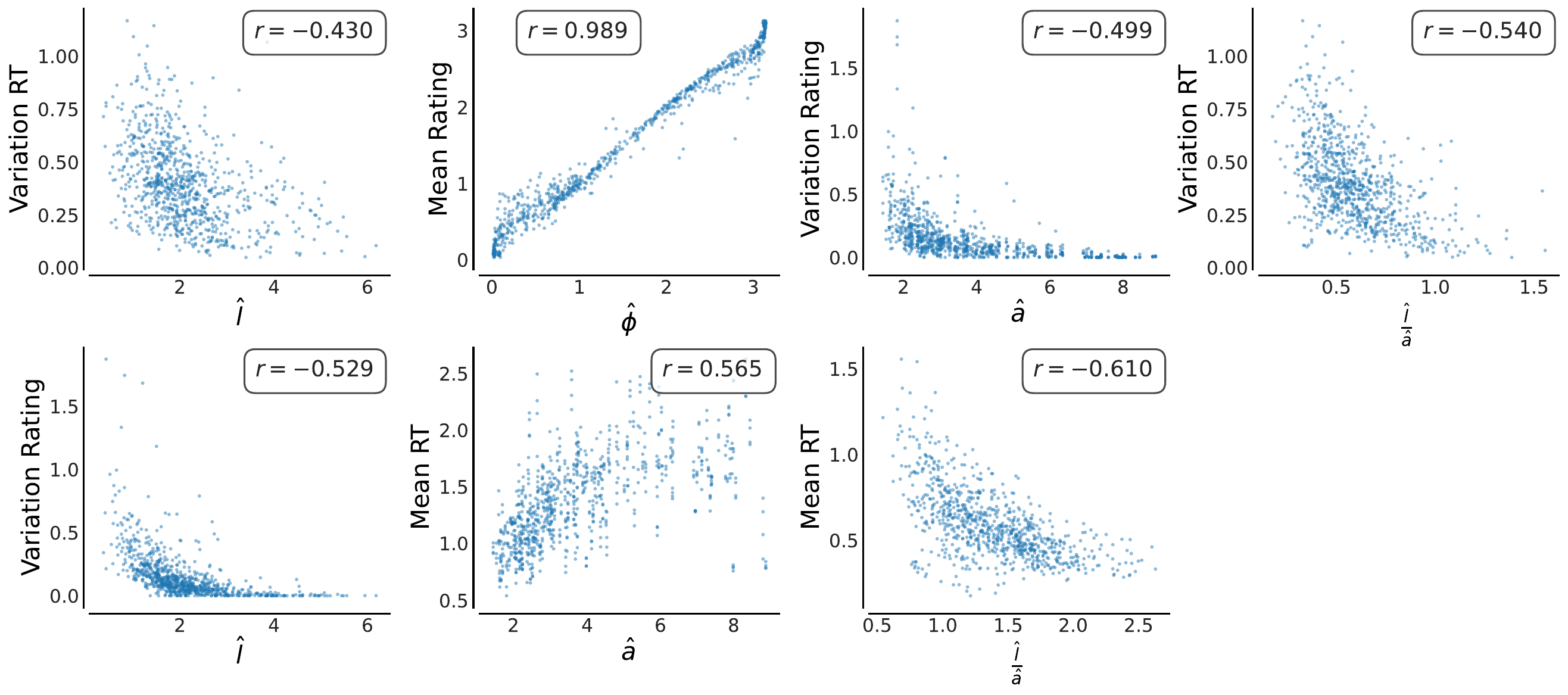}
 \caption{Correlation between HCDM parameters and empirical summary statistics. Each subplot shows one parameter–statistic pair, with the x-axis representing the estimated parameter and the y-axis representing the corresponding summary statistic. Data from all participants are included.} 
 \label{fig:HCDMcor}
\end{figure}

To validate the interpretation of the model parameters, we computed Pearson correlations between the estimated parameters and the summary statistics of the empirical data (see Table~\ref{tab:HCDMcor} and Figure~\ref{fig:HCDMcor}). The drift rate length $l$ in HCDM correlates negatively with the variation in reaction time ($r = -0.430$) and the variation in ratings ($r = -0.529$), from which it can be concluded that faster accumulation leads to less variability in both the time taken and the final rating. The $\phi$ showed a strong linear positive correlation with the mean affective ratings ($r = 0.989$), reflecting the direction of the evidence accumulation. The specific interpretation of $\phi$ is task-dependent. Given that the rating scale in this dataset represents affect intensity (from negative to positive), $\phi$ thus simply captures the average affect intensity within a condition. The decision boundary parameter $a$ has a positive correlation with the mean reaction time ($r = 0.565$) and a negative correlation with the variation in ratings ($r = -0.499$). This aligns with the common interpretation that more cautious participants (higher $a$) take longer to decide and respond more consistently on the rating scale. These relationships also align with what has been found analytically in the CDM \citep{smith2016diffusion, rogers2000diffusions}, where the theoretical distribution of the responses is the von Mises distribution, and the precision of the von Mises distribution is $l \times a$ (assuming that the diffusion coefficient unit is 1). Lastly, we computed the ratio $\frac{l}{a}$, which is the relative evidence accumulation speed. This ratio negatively correlates with the mean reaction time ($r = -0.610$), since easier decisions are faster reactions. The ratio also negatively correlates with the variation in reaction time ($r = -0.540$), because when the relative speed is high, the efficient evidence accumulation means it is much less likely that the noises will meander for an extended period before a decision is made, resulting in a lower variation. Overall, the model parameters are identifiable, and the fitted parameters can be directly mapped to distinct psychological processes such as caution, processing speed, decision efficiency, and affective intensity.

\begin{table}[h]
\centering
\begin{tabular}{crcccc}
\toprule
HCDM parameters & $\text{rt}_{mean}$ & $\text{rt}_{var}$ & $\text{rating}_{mean}$ & $\text{rating}_{var}$  \\
\midrule
$l$           & 0.178 & -0.430 & 0.013 & -0.529 \\
$\phi$      & -0.015 & 0.007  & 0.989 & -0.105 \\
$a$           & 0.565  & -0.075 & 0.016 & -0.499  \\
$z_1$           & 0.047  & 0.087 & -0.188 & 0.098  \\
$z_2$           & -0.227  & -0.055 & 0.004 & -0.159  \\
$T_{er}$       & -0.179  & -0.036 & -0.012 & 0.308  \\
$\frac{l}{a}$ & -0.610 & -0.540 & 0.035 & -0.009 \\
\bottomrule
\end{tabular}
\caption{Correlations between the HCDM parameters and the summary statistics of the empirical data.}
\label{tab:HCDMcor}
\end{table}

\newpage

\subsubsection*{BDDM}

As in the HCDM, we assumed that four outcomes led to four drift rate functions, while the other parameters remained constant across conditions. 

First, parameters were sampled from their respective priors. For the drift rate function, we sampled $\mu \sim U(0.15, 0.85)$ and $h \sim U(0.2, 3)$. The sampling distribution of $\kappa$ depended on the sampled $\mu$: $\kappa \sim U\big(\max(\frac{1}{\mu}, \frac{1}{1-\mu}), 20\big)$. This constraint prevents the Beta density from exhibiting a U-shape (with infinite density near 0 or 1), which would be unrealistic as a drift rate function. The remaining parameters were sampled as $a \sim U(0.5, 9)$, $T_{er} \sim U(0.1, 2.5)$, and $\rho \sim U(0.01, 0.5)$. We then simulated data sets with choice reaction times based on the aforementioned priors. The simulations were performed using the Euler–Maruyama method (see Equation~\ref{eq:bddm}) with $\Delta t = 0.001 \, s$. The trial number per condition was sampled uniformly from the integers 39 through 50, matching the empirical data.

We simulated 100,000 datasets with known ground-truth parameters to train the neural networks; generating these datasets required approximately 10 minutes on an NVIDIA Tesla H100-SXM2-32GB GPU. The summary network has the same architecture as the HCDM estimator. The inference network is a conditional flow matching network with a 3-layer MLP (512 units per layer). The neural networks were trained for 100 epochs (batch-size 256, offline training) using the Adam optimizer (initial learning rate $5 \times 10^{-4}$, a cosine decay schedule), and loss had converged by the end of training. Each epoch took approximately 4 seconds, resulting in a total training time of 6.67 minutes. Importantly, we used offline training: all 100,000 simulated datasets were generated before training began and were then repeatedly reused throughout optimization. This contrasts with online training, in which new datasets are simulated at every optimization step and never reused. We adopted the offline approach because simulating data from the BDDM is computationally expensive and constitutes the primary computational bottleneck, making offline training more suitable for simulation-expensive models.

\begin{figure}[!ht] 
 \centering
 \includegraphics[width=1\linewidth]{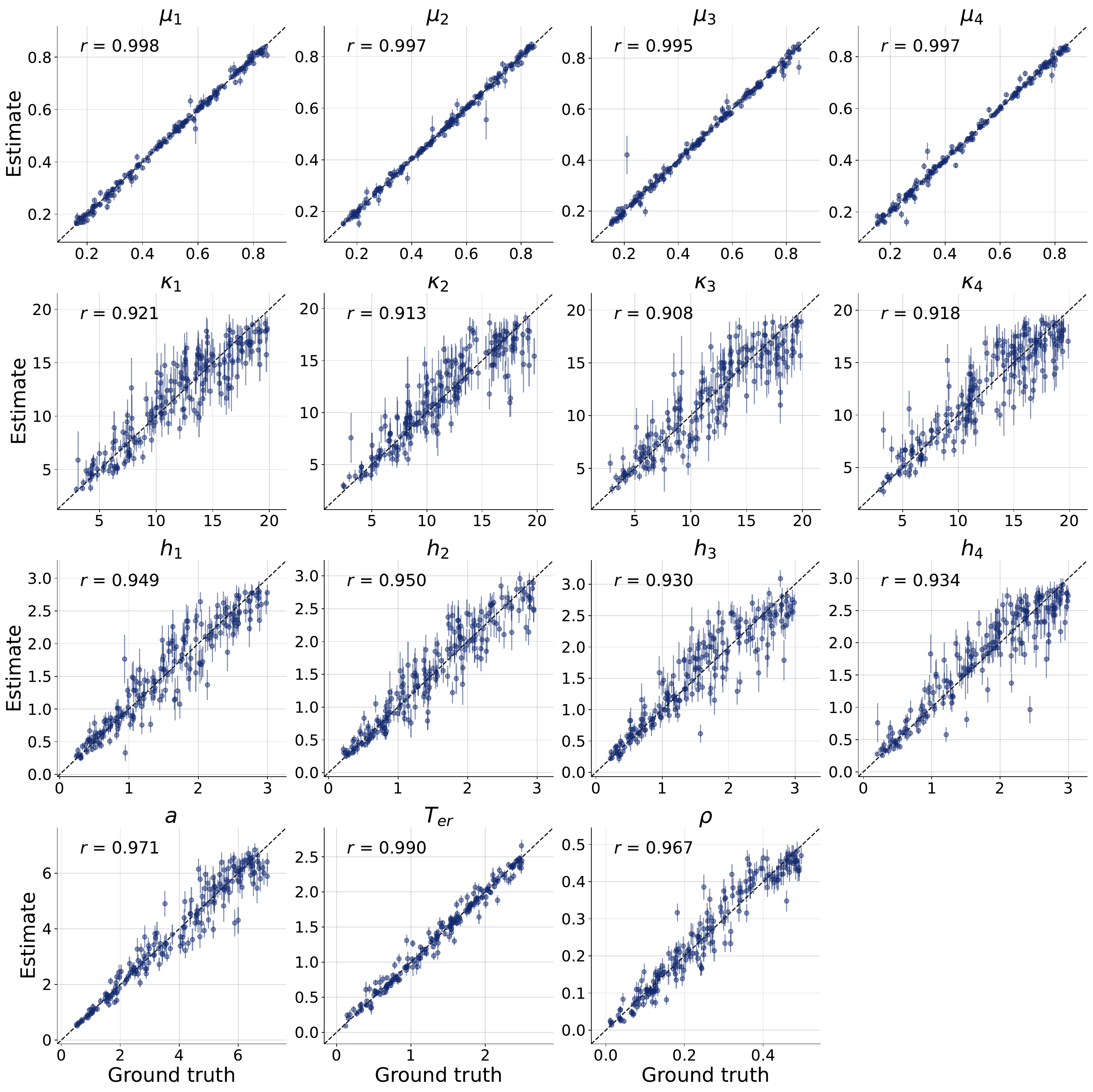}
 \caption{The parameter recovery of the BDDM. The x-axis stands for the ground-truth parameter values, and the y-axis represents their estimates. The black dashed line is the identity line (indicating perfect recovery). In the top left corner the correlation between true and the estimated values is shown.} 
 \label{fig:BDDMrecovery}
\end{figure}

We further simulated 200 validation datasets to assess parameter recovery (see Figure~\ref{fig:BDDMrecovery}) and found that the parameters were generally well recovered. With a limited number of trials, $\mu$ is recovered more accurately and with lower uncertainty than the concentration parameter $\kappa$. Overall, parameter uncertainty was greater than in the HCDM, partly due to the increased complexity of the BDDM (e.g., the drift rate in the HCDM is parameterized by two parameters, whereas in the BDDM it is parameterized by three). Meanwhile,  the simulation-based calibration shows that the estimates were unbiased (see Figure~\ref{fig:ecdf_bddm}).

\begin{figure}[!ht] 
 \centering
 \includegraphics[width=1\linewidth]{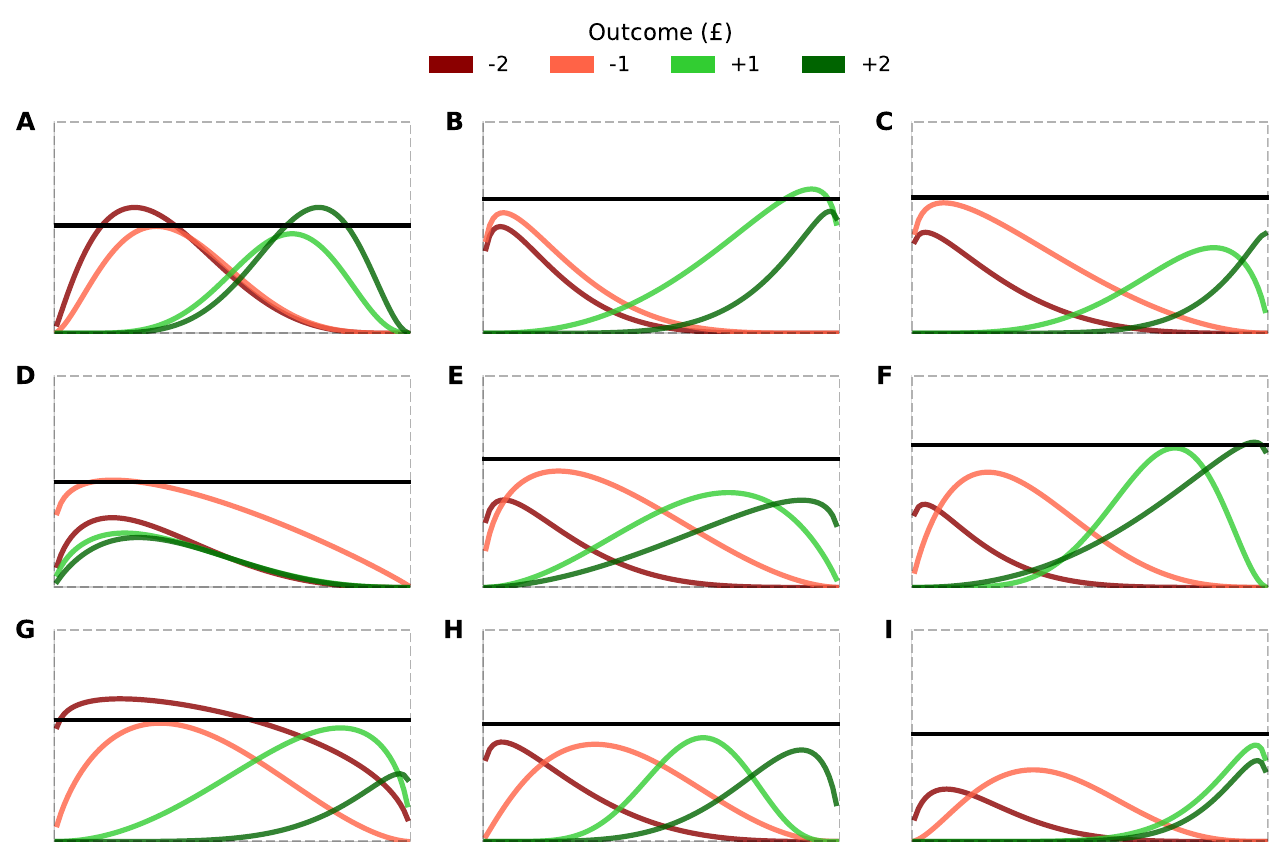}
 \caption{Visualization of BDDM estimates. We randomly sampled nine participants, extracted the estimated posterior mean, and plotted the Beta-shaped drift rate functions (the colored curves) and the decision boundary (the black horizontal line).} 
 \label{fig:BDDMfitting}
\end{figure}

We then fitted the empirical data to the model by inputting each participant's data into the trained neural networks to obtain posterior samples. Posterior sampling for all 215 participants took approximately 664 milliseconds (1000 samples per participant). Figure~\ref{fig:BDDMfitting} shows the estimated mean of drift rate functions (the colored curves) and the decision boundary (the horizontal line) for the same nine participants as in Figure~\ref{fig:HCDMfitting} \footnote{Similar to Figure~\ref{fig:HCDMfitting}, the drift rates cross the boundary in some subplots (participants A, B, D, F, and G). This does not imply that decisions are made in the first time step; rather, the average evidence accumulated along the line is the drift rate function times $\Delta t$. }. Except for participant D, participants generally showed a clear relationship between outcomes and drift rates function: Lower outcomes were associated with drift rates with smaller $\modedrift$, indicating lower affect ratings. Additionally, the more extreme the absolute value of the outcome, the more concentrated the drift rate.

\begin{figure}[!ht] 
 \centering
 \includegraphics[width=1\linewidth]{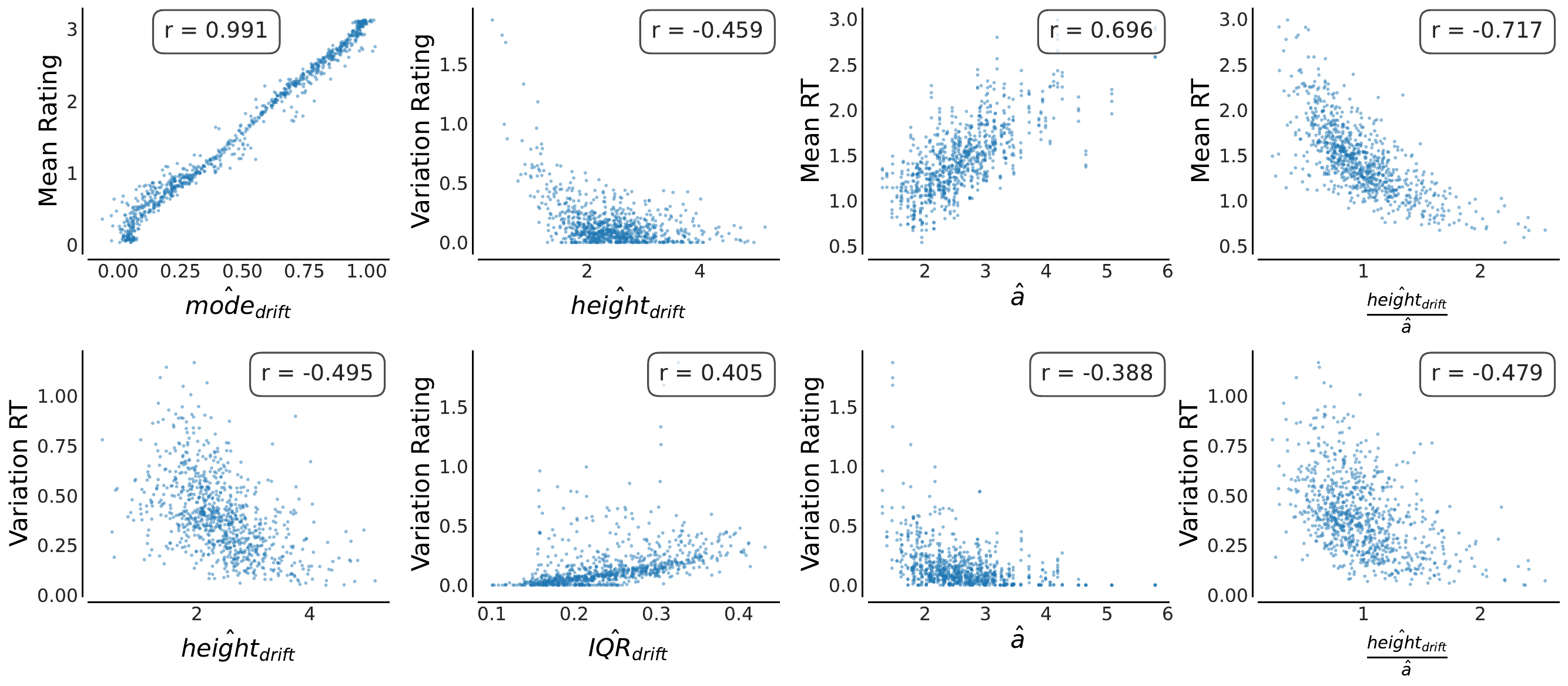}
 \caption{Correlation coefficients between BDDM parameters/ features and empirical summary statistics. Each subplot shows one parameter–statistic pair, with the x-axis representing the estimated parameter and the y-axis representing the corresponding summary statistic. Data from all participants are included.} 
 \label{fig:BDDMcor}
\end{figure}

To validate the interpretation of BDDM parameters, we computed Pearson correlations between the parameters and summary statistics of the empirical data (see Table~\ref{tab:BDDMcor} for full results and Figure~\ref{fig:BDDMcor} for visualizations of selected parameter–statistic pairs). As expected, $\modedrift$ was highly correlated with the mean ratings ($r = 0.991$), which is the direction of evidence accumulation. $\IQRdrift$ showed a positive correlation with the variation in ratings ($r = 0.405$), confirming that more concentrated drift rates lead to more concentrated ratings. The interpretation of this parameter depends on the task, but generally $\IQRdrift$ reflects the clarity of evidence accumulation.  

Additionally, $\heightdrift$ correlated negatively with the variation in ratings ($r = -0.459$), suggesting that faster evidence accumulation leads to less variability in the final decisions, similar to the effect of the drift rate length $l$ in the HCDM. The decision boundary $a$ was positively correlated with the mean reaction time ($r = 0.696$) and negatively correlated with the variation in ratings ($r = -0.388$). These effects align with those observed in HCDM (see Table~\ref{tab:HCDMcor} and Figure~\ref{fig:HCDMcor}) and are consistent with the original SCDM \citep{ratcliff2018decision}, in which increasing the decision boundary led to a modest decrease in response variance and a large increase in mean RT in simulated data. 

We also computed the ratio of $\heightdrift$ to $a$, representing the fastest evidence accumulation speed relative to the decision boundary. Consistent with what we found with HCDM, this ratio was negatively correlated with both the mean reaction time ($r = -0.717$) and the variation in reaction time ($r = -0.479$). 

Taken together, both HCDM and BDDM have parameters that are meaningful and correlated to the empirical summary statistics. As the drift rate is of primary interest in EAMs, we present a summary table (Table~\ref{tab:interpretation}) to illustrate the mapping of drift rate to evidence accumulation features across the three models.

\begin{table}[!ht]
\centering
\begin{tabular}{crcccc}
\toprule
BDDM parameters/ features & $\text{rt}_{mean}$ & $\text{rt}_{var}$ & $\text{rating}_{mean}$ & $\text{rating}_{var}$  \\
\midrule
$h$                      & -0.198 & -0.243 & -0.247 & -0.028  \\
$\mu$                    & -0.025 & 0.002  & 0.984  & -0.105 \\
$\kappa$                 & -0.067 & -0.109 & 0.119  & -0.200 \\
$\text{height}_\text{drift}$    & -0.306  & -0.120 & 0.022 & -0.459 \\
$\text{mode}_\text{drift}$      & -0.021  & 0.002 & 0.991 & -0.103 \\
$\text{IQR}_\text{drift}$       & 0.046  & 0.133 & -0.222 & 0.405 \\
$a$                             & 0.696 & 0.088 & -0.022  & -0.388 \\
$T_{er}$                        & -0.046 & -0.163  & -0.014  & 0.047 \\
$\rho$                         & 0.025 & 0.053  & -0.005  & -0.133 \\
$\frac{\text{height}_\text{drift}}{a}$& -0.717 & -0.479 & -0.123 & -0.144 \\

\bottomrule
\end{tabular}
\caption{Correlation coefficients between BDDM parameters and empirical summary statistics.}
\label{tab:BDDMcor}
\end{table}

\begin{table}[!ht]
\centering
\begin{tabular}{cllc}
\toprule
Model & Drift rate & Mapping to evidence accumulation \\
\midrule
    DDM     & $|v|$       & Speed \\
            & $+/-$       & Direction (binary) \\
\midrule
    HCDM    &  $l$      & Speed \\
            &  $\phi$ & Direction (bounded, continuous) \\
\midrule
    BDDM    &  $\text{height}_\text{drift}$ & Speed\\
            &  $\text{mode}_\text{drift}$ & Direction (bounded, continuous)\\
            & $\text{IQR}_\text{drift}$ & Evidence clarity\\
\bottomrule
\end{tabular}
\caption{Mapping of drift rate to evidence accumulation features in three EAMs: the regular drift diffusion model (DDM), the half-circular diffusion model (HCDM), and the beta drift diffusion model (BDDM).}
\label{tab:interpretation}
\end{table}

\newpage
\subsection*{Model check}
After fitting the models, we conducted posterior predictive checks for both models to evaluate how well each model could reproduce the observed data \citep{berkhof2000posterior, gelman2013bayesian}. The procedure was as follows: (a) We obtained 2{,}000 posterior draws for each participant; (b) from these, we randomly sampled 100 posterior draws $\btheta_{(ik)}$ per participant $i$ to account for posterior uncertainty; (c) for each $\btheta_{(ik)}$, we generated a synthetic dataset of the same size as the origin, yielding 100 replicates per participant; (d) to summarize the continuous RT and rating distributions, we calculated the 0.1, 0.3, 0.5, 0.7, and 0.9 quantiles for every simulated dataset; (e) we averaged the predicted quantiles across the 100 simulated datasets for each participant. These average predictions were then plotted against the empirical quantiles to visualize the models' ability to recover the central tendency, spread, and skewness of the observed data.

\begin{figure}[!ht] 
 \centering
 \includegraphics[width=1\linewidth]{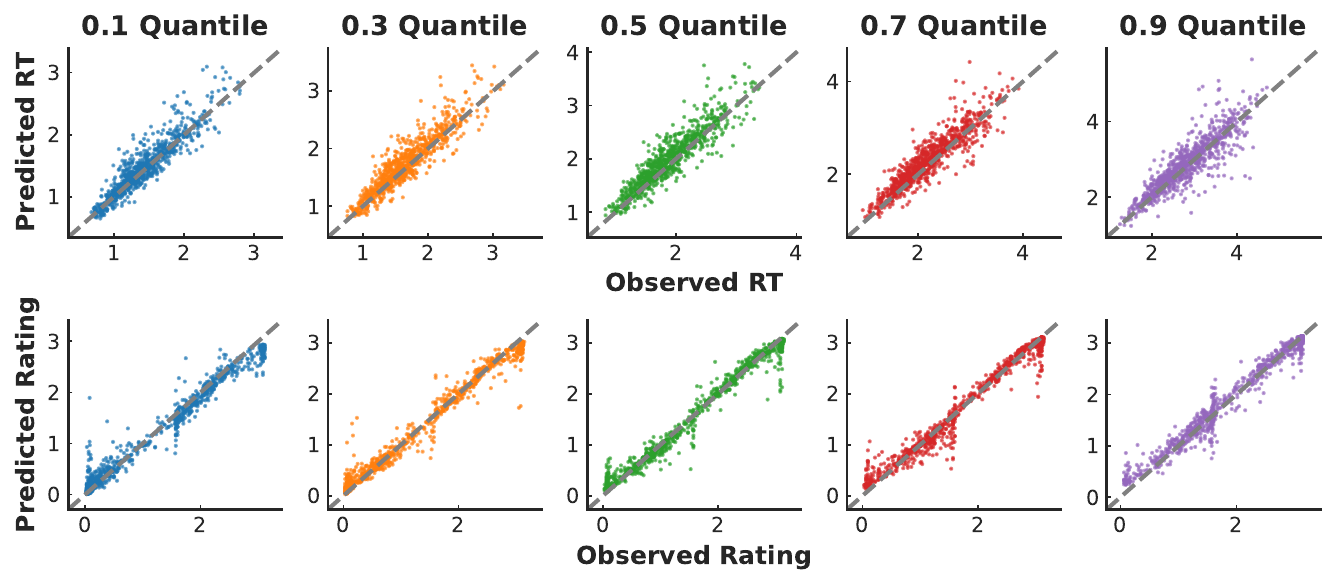}
 \caption{The posterior predictive check of HCDM. The first row shows the observed quantiles of reaction time per condition from each participant against their predicted values. The second row shows the observed quantile of responses against their predicted value. Note that the range of the ratings was rescaled from 0 to $\pi$ and remains continuous.} 
 \label{fig:HCDMppc}
\end{figure}

\begin{figure}[!ht] 
 \centering
 \includegraphics[width=1\linewidth]{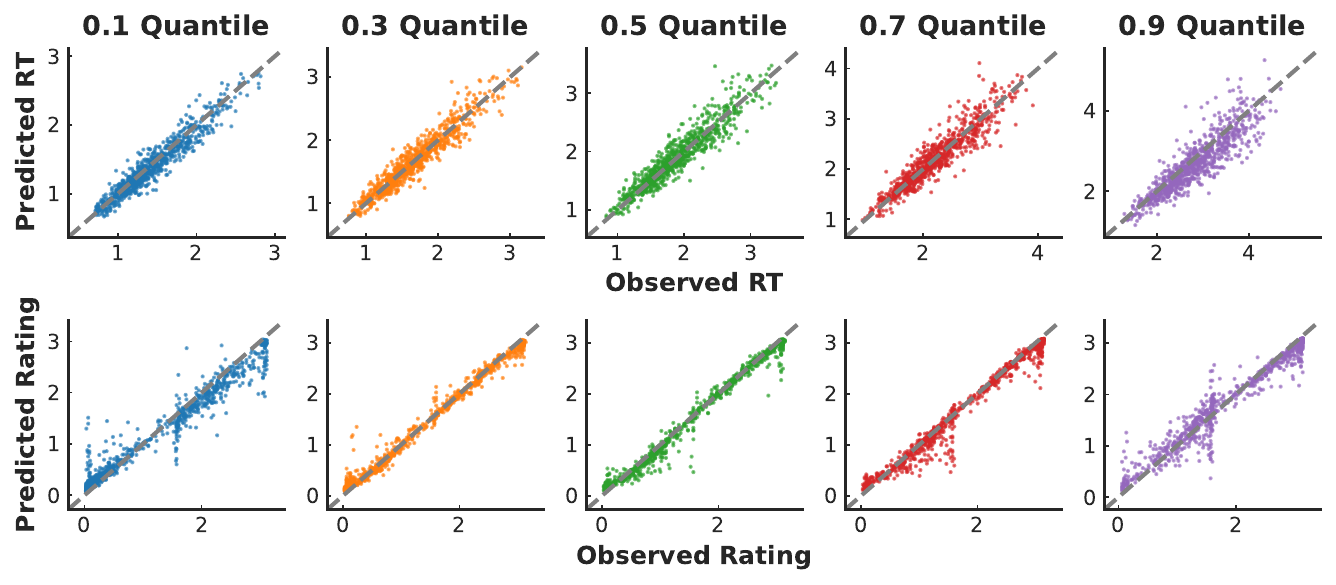}
 \caption{The posterior predictive check of BDDM. The first row shows the observed quantiles of reaction time per condition from each participant against their predicted values. The second row shows the observed quantile of responses against their predicted value. Note that the range of the ratings was rescaled from 0 to $\pi$ and remains continuous.} 
 \label{fig:BDDMppc}
\end{figure}

As shown in Figures~\ref{fig:HCDMppc} and~\ref{fig:BDDMppc}, both models produced predictions that closely matched the observed data, particularly for the reaction-time quantiles. However, both models exhibited slight systematic biases in predicting rating quantiles: they tended to overestimate quantiles when the observed quantiles were near the left end of the scale and underestimate them when near the right end. Additionally, both models produced several outlying predictions when the observed quantiles are near the midpoint of the rating scale.

\begin{figure}[!ht] 
 \centering
 \includegraphics[width=1\linewidth]{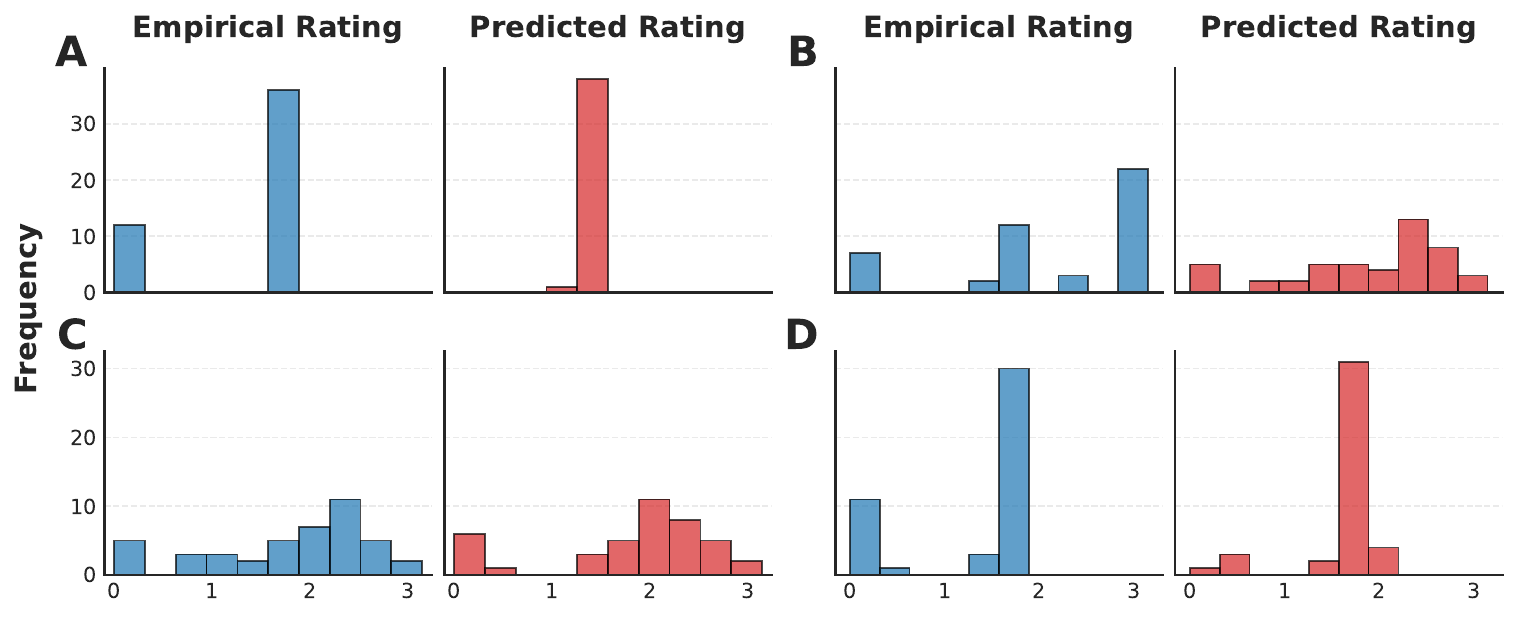}
 \caption{Overestimation of the 0.1 quantile of rating using BDDM. Panels A-D show four different participants. The empirical and predicted ratings shown are specifically from the condition where the model failed to accurately predict its 0.1 quantile (i.e., when the deviation exceeded $0.8$).} 
 \label{fig:BDDMprobrating}
\end{figure}

To further examine the overestimation of rating quantiles near zero, we inspected four cases in which the deviation exceeded 0.8 (see Figure~\ref{fig:BDDMprobrating}). In these cases, the empirical rating distributions were highly clustered and multimodal, with a pronounced bump around zero that the model failed to capture. We observed a highly similar pattern for the outlying predictions when the observed rating quantiles near the midpoint of the scale (see Figure~\ref{fig:BDDMprobrating2}): The models struggled to account for rating distributions that were both sparse and strongly clustered. Such patterns arise when participants tend to select extreme values or remain anchored at specific points on the scale (e.g., the “neutral” point). In contrast, the predicted reaction-time quantiles exhibited less systematic bias, likely because reaction times follow a uni-modal continuous distribution.

Overall, aside from these failures in capturing atypical clustered multimodal rating patterns, both models predicted the distributions of empirical data well.

\newpage

\subsection*{Model comparison}

\subsubsection*{Method}

Besides model checking, it is informative to compare both models directly and perform model comparison (or model selection). Similar as in the model fitting, here we used the amortized methods to learn posterior model probabilities (PMP; i.e., the probability of each model given the data). For each participant, we compared two competing models. As a generic notation, we refer to these models as $\mathcal{M}_1$ (the HCDM) and $\mathcal{M}_2$ (the BDDM). Although we only considered two models, the explanation below will be given for the general case of $J$ models. Given our observed data $\bx_\text{obs}$, the inference target is the posterior model probabilities (PMPs):
\begin{equation}
    p(\mathcal{M}_j \mid \bx_\text{obs}) = \frac{p(\bx_\text{obs} \mid \mathcal{M}_j)p(\mathcal{M}_j)}{\sum_k p(\bx_\text{obs} \mid \mathcal{M}_k)p(\mathcal{M}_k)}
\end{equation}
where $p(\mathcal{M}_j)$ is the prior probability for model $j$ (with $j=1,\dots,J$). 

Computing these PMPs is a challenging task because they depend on the marginal likelihood or evidence $p(\bx_\text{obs} \mid \mathcal{M}_j) = \int p(\bx_\text{obs} \mid \btheta, \mathcal{M}_j) p(\btheta| \mathcal{M}_j) d\btheta$, which involves an integral over the parameter space. To circumvent this computation, we frame the task as a probabilistic classification for which the neural networks are trained to learn the mapping between data and the models PMPs. The neural network consists of a summary network (learning summary statistics), and an evidential neural network that outputs PMPs. This approach is known as amortized Bayesian model comparison (ABMC). For the details of ABMC, see the Appendix.

ABMC is known to be sensitive to model misspecification \citep{elsemuller2023sensitivity}, that is, if none of the models represent the data-generating process for the observed data, the networks will produce unstable predictions \citep{cannon2022investigating}. To avoid that, we applied deep ensembles \citep{elsemuller2023sensitivity}, where $k$ identically configured but independently initialized neural networks were trained. This allowed us to average their PMPs and use the standard deviation of across-network PMPs as a measure of uncertainty. Furthermore, we simulated contaminants in the training data by assuming each data point has 10\% chance to be contaminated. The contaminated reaction time came from $U(0.1, 5)$, while the contaminated responses were drawn from $U(0, \pi)$. This is a standard way of increasing the robustness of approximators when using amortized simulation-based methods \citep{wu2026testing}. By using deep ensembles and simulating contaminants in training data, we obtained more reliable model comparison results. 

Additionally, we present a comparison of the average PMPs from ABMC and the difference in goodness of fit ($\Delta \mathrm{GOF}$) from the posterior predictive checks. This analysis serves as an additional validation of ABMC. 

To calculate $\Delta \mathrm{GOF}$ from the posterior predictive checks, we computed both observed quantiles $Q^\text{obs}$(0.1, 0.3, 0,5, 0.7, 0.9) and predicted quantiles $\hat{Q}$ from both the HCDM and the BDDM. For each participant, model fit can be quantified as the total absolute deviation between observed and predicted quantiles across five quantile levels indexed by $i$ and four conditions indexed by $j$: 
\begin{equation}
    \mathrm{GOF} = \sum_{i=1}^{5}\sum_{j=1}^{4}|\hat{Q}_{ij} - Q^{obs}_{ij}|. 
\end{equation}
This quantity was computed separately for reaction times ($\mathrm{GOF}_{rt}$, ranging from 0.1 to 5) and responses ($\mathrm{GOF}_{resp}$, ranging from 0 to $\pi$). To obtain a combined goodness-of-fit measure, we simply calculated a weighted sum: For each participant, we added up the $\mathrm{GOF}_{rt}$ (ranging from 0.1 to 5) and $\mathrm{GOF}_{resp}$ (ranging from 0 to $\pi$):
\begin{equation}
    \mathrm{GOF}_{total} = \mathrm{GOF}_{rt} + \frac{5}{\pi} \mathrm{GOF}_{resp},
\end{equation}
where the scaling factor ensured that the two components were on a comparable scale. The $\Delta \mathrm{GOF}$ was defined as $\mathrm{GOF}^\text{BDDM}_{total} - \mathrm{GOF}^\text{HCDM}_{total}$ such that negative values indicated better predictive performance of the BDDM relative to the HCDM.

\subsubsection*{Implementation and results}

To perform model comparison, we trained $k = 10$ identical PMP approximators. We simulated datasets from both the HCDM and the BDDM, with each model being equally likely. The simulation of datasets was the same as described in the model fitting section. The only difference was that each dataset now had a model label attached to it (e.g., 1 $=$ HCDM and 2 $=$ BDDM). 200,000 datasets were simulated beforehand to train the approximators. Each approximator used a SetTransformer \citep{lee2019set} with 128-dimensional outputs as the summary network and a simple multilayer perceptron as the evidential network (4 layers with 256 units per layer and elu activations). Training was performed for 30 epochs (batch size = 256, offline) using the Adam optimizer with a cosine decay learning-rate schedule (initial learning rate = $5 \times 10^{-4}$). Training each approximator took approximately 4 minutes on an NVIDIA Tesla H100-SXM2-32GB GPU. We evaluated each approximator using calibration curves \citep{elsemuller2024deep} and confusion matrices, confirming accurate predictions across all networks (see Figures~\ref{fig:mccalibration} and~\ref{fig:confusionmatrix} for representative examples).

\begin{figure}[!ht] 
 \centering
 \includegraphics[width=1\linewidth]{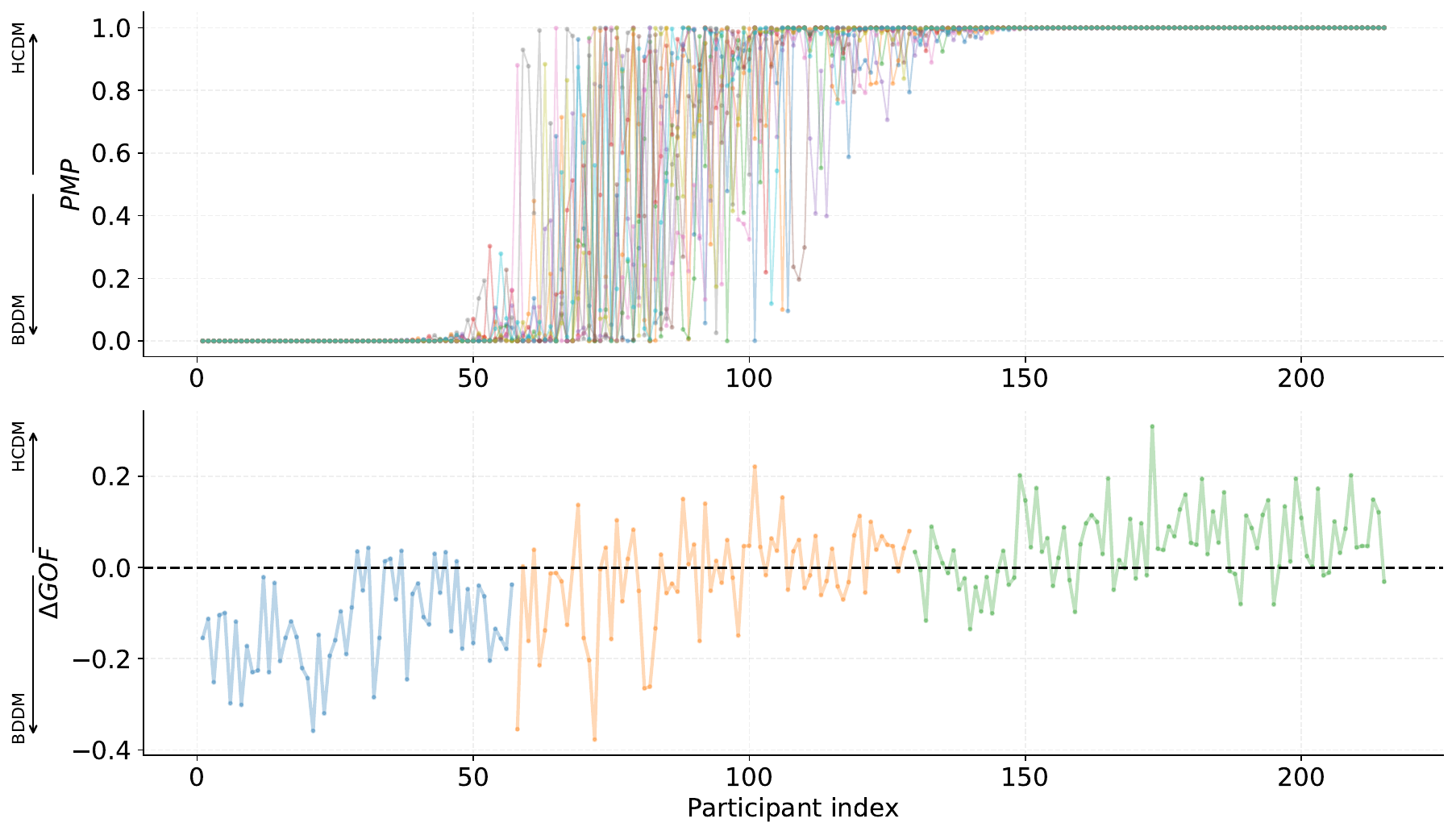}
 \caption{The $\text{PMP}_\text{HCDM}$ and $\Delta \mathrm{GOF}$. The top panel shows the $\text{PMP}_\text{HCDM}$ for each participant, ordered by their average PMP across 10 independently trained approximators. Participants on the left strongly favor the BDDM, those on the right strongly favor the HCDM, and those in the middle show mixed evidence. The lower panel shows $\Delta \mathrm{GOF}$ ordered accordingly, with blue, yellow, and green segments indicating participants who strongly favor the BDDM, show mixed evidence, and strongly favor the HCDM, respectively. } 
 \label{fig:pmp}
\end{figure}

Next, we input the empirical data into the 10 approximators to obtain PMPs conditioned on the data from 215 participants. The estimated PMPs for the HCDM are shown in the upper panel in Figure~\ref{fig:pmp}. Because only two models were compared, the PMP of the alternative model (BDDM) is simply $1 - \text{PMP}_{\text{HCDM}}$.
The results indicate that around 57 participants strongly favored the BDDM, with highly consistent PMPs across approximators, suggesting no evidence of model misspecification. About 90 participants showed strong evidence in favor of the HCDM. The remaining participants, as shown in the middle of Figure~\ref{fig:pmp} (upper panel), exhibited moderately to highly mixed evidence across models. In the lower panel of Figure~\ref{fig:pmp}, we plot the corresponding $\Delta \mathrm{GOF}$ values. The blue line represents participants who strongly favored the BDDM (with an average $\text{PMP}_\text{HCDM} < 0.95$), with $\Delta \mathrm{GOF}$ generally below zero. The yellow line corresponds to participants with mixed PMPs, whose $\Delta \mathrm{GOF}$ values were around zero. The green line represents participants who strongly favored the HCDM (with an average $\text{PMP}_\text{HCDM} > 0.95$) and whose $\Delta \mathrm{GOF}$ values were generally above zero. These patterns indicate that the model assignments obtained from the neural network–based PMPs are consistent with the relative goodness of fit measured in the posterior predictive checks. Overall, more participants showed strong evidence for the HCDM than for the BDDM.

\begin{figure}
    \centering
    \includegraphics[width=1\linewidth]{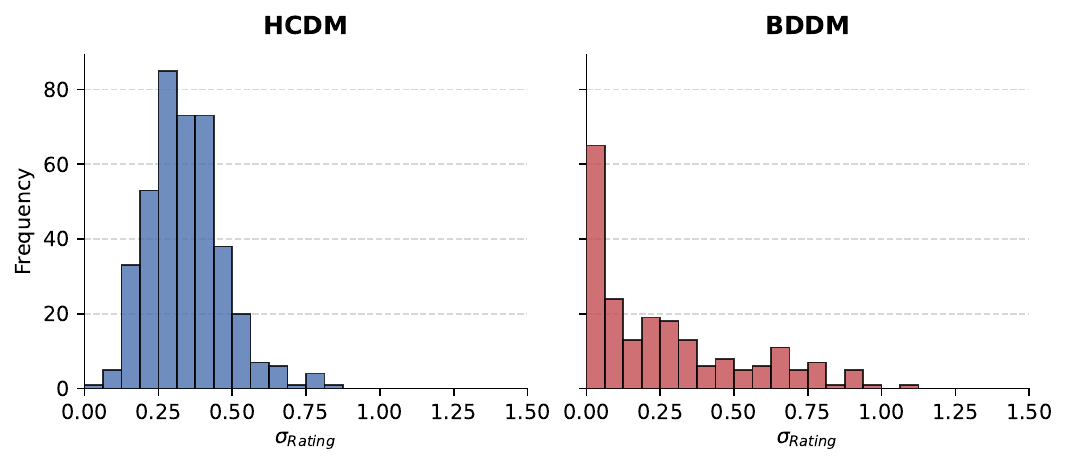}
        \caption{Histogram of rating standard deviations. The left panel shows datasets best explained by the HCDM, and the right panel shows datasets best explained by the BDDM. Standard deviations were computed from ratings within each condition for each dataset.}
    \label{fig:pmpsd}
\end{figure}

We extracted the data from participants who had a predicted $\text{PMP}_\text{HCDM}$ or a predicted $\text{PMP}_\text{BDDM}$ larger than 90\%, and examined differences in the rating distributions between the two groups. As shown in Figure~\ref{fig:pmpsd}, the most salient difference was how concentrated the ratings were: datasets favoring the HCDM showed moderate spread (SD ranging from approximately 0.10 to 0.50), whereas datasets favoring the BDDM were either highly concentrated or highly dispersed (SD smaller than 0.02 or up to 1.00). This pattern reflects a fundamental difference in model flexibility: the BDDM includes a kernel parameter $\rho$, governing Gaussian process noise, which can produce either very jagged or nearly flat trajectories, allowing it to fit both highly concentrated and highly dispersed rating distributions. The HCDM, lacking this mechanism, predominantly fits data with moderate spread. Consequently, the degree of concentration in the rating distribution is a key determinant of which model better captures the data.

\subsection*{Discussion}

Bounded continuous self-report data are among the most widely collected forms of behavioral data in psychology, yet they have remained largely outside the reach of the EAM framework. In this paper, we addressed this gap by formulating and validating two stochastic EAMs designed for such data: the HCDM, which restricts the CDM's circular decision space to a semi-disc via a reflecting boundary, and the BDDM, which replaces the SCDM's stimulus-anchored Gaussian drift rate with a freely estimated scaled beta density. Using ABI and ABMC, we showed that both models have identifiable parameters that can be reliably recovered, capture the joint distribution of ratings and reaction times in an empirical affect dataset, and yield interpretable mappings between parameters and observable behavior. For instance, larger decision boundaries are associated with longer mean reaction times and lower rating variability in both models. We organize the remainder of the discussion around four questions: how these models fit into the broader EAM family, when each should be used, what they imply for the cognitive interpretation of self-report, and what they cannot yet do.

\paragraph{Positioning within the EAM framework.} HCDM and BDDM complete a natural progression within the EAM family. The DDM handles binary choice; the CDM and SCDM handle rotationally symmetric or stimulus-anchored continuous responses; HCDM and BDDM handle bounded continuous responses without an externally specified target. Each step preserves the core EAM commitment — joint modeling of choice and reaction time via stochastic evidence accumulation — while adapting the geometry of the decision space and the parameterization of the drift rate to the response domain. A parallel line of work has explored deterministic rather than stochastic accumulation for continuous responses \citep{kvam2023unified}; our models occupy the stochastic branch, preserving the trial-to-trial variability that has been central to EAM accounts of cognition. Within the stochastic branch, HCDM and BDDM represent two complementary parameterizations: HCDM inherits the geometric simplicity of the CDM (drift as a vector, noise as standard Brownian motion), while BDDM inherits the SCDM's richer notion of a spatially distributed drift rate together with spatially correlated noise. As we show below, this difference in flexibility has direct empirical consequences for which model fits which data.

\paragraph{Practical guidance: which model to use?} Our model comparison results suggest that the choice between HCDM and BDDM depends primarily on the dispersion structure of the rating data. The HCDM provides a parsimonious account for data with moderate rating spread (within-condition SDs roughly between 0.10 and 0.50 on the rescaled $[0, \pi]$ interval) and is computationally cheaper to fit because of its simpler parameterization and lower-dimensional noise. The BDDM, with its Gaussian-process noise kernel $\rho$, accommodates both highly concentrated and highly dispersed ratings, at the cost of an additional parameter and substantially more expensive simulation (roughly an order of magnitude longer in our setup). As a default, we recommend fitting both models in parallel and using ABMC to identify the better-supported model per participant; when computational resources are limited, the HCDM is a reasonable single-model choice unless the data exhibit visibly bimodal or near-degenerate rating distributions. Neither model is well-suited to data with strong anchoring effects (e.g., heavy mass at scale endpoints or anchored midpoints), and researchers encountering such patterns should consider mixture extensions or experimental modifications discussed below.

\paragraph{Theoretical implications.} Beyond their methodological contribution, HCDM and BDDM open a path toward formal cognitive accounts of self-report. Self-report is often treated as a transparent readout of an internal state, but a growing literature reframes it as a decision process in its own right \citep{teoh_framing_2023, givon_are_2023}. Our models make this framing operational on bounded continuous scales: the drift rate captures the speed, direction and clarity of the internal evidence being reported, the boundary captures response caution, and the non-decision time captures perceptual and motor components separable from the deliberation itself. In our application, this decomposition revealed individual differences that would be invisible to summary-statistic analyses. For example, participants whose ratings differed primarily in caution versus those whose ratings differed in evidence clarity would produce similar marginal rating distributions but distinct joint distributions with reaction time. This kind of process-level resolution is the central reason EAMs have been productive in two-alternative choice research, and our work makes the same resolution available for bounded continuous self-report.

\paragraph{Limitations.} Several limitations warrant explicit acknowledgment. First, both models assume a continuous response process and therefore fail to capture clustered or multimodal rating patterns that arise when participants anchor on specific scale points (e.g., ``neutral''). In our dataset this affected a small but non-trivial subset of participants, and posterior predictive checks revealed systematic biases near the scale endpoints. Future experimental designs may benefit from reducing or removing scale anchors, or alternatively, future models could incorporate explicit mixture components for anchored responding. Second, the BDDM is computationally expensive: its Gaussian-process noise requires offline training on pre-simulated datasets, and adding parameters (e.g., trial-to-trial variability in drift or boundary) would compound this cost. Third, our ABMC procedure compares two specific model variants and does not exhaustively explore the space of plausible bounded-continuous EAMs; further alternatives, such as a half-circular model with an additional flexibility parameter, or a beta model with a simpler noise structure, could refine the parameter choices we made. Fourth, the empirical application uses a single dataset from a probabilistic reward paradigm; generalization to other self-report constructs (confidence, risk preference, attitude reports) remains to be tested.

\paragraph{Future directions.} Several directions follow from this work. First, extending the models to accommodate discrete or anchored response strategies (e.g., through mixture components for continuous and categorical responding, analogous to ZOIB models in item response theory \citep{molenaar2022zero}) could improve their ability to capture clustered and multimodal distributions. Second, linking model parameters to experimental manipulations or neural measures could strengthen their psychological interpretation and enable hypothesis tests about which latent process is affected by a given manipulation. Third, the ABI and ABMC frameworks make it feasible to explore richer model spaces, including hierarchical extensions that pool information across participants, hybrid models combining stochastic accumulation with deterministic components \citep{kvam2023unified}, and adaptive evidence-accumulation mechanisms in which drift parameters evolve within a trial. Finally, applying these models to longitudinal self-report data — for example, ecological momentary assessment of affect — could reveal how decision-process parameters themselves vary over time and context, opening a new window onto the dynamics of subjective experience.

\section*{Declaration}
\subsection*{Funding}
Yufei Wu, Tamás Szűcs, Agnes Moors, and Francis Tuerlinckx were partially supported by a grant from the Research Council of KU Leuven (C14/23/062).

\subsection*{Conflicts of interest} 
The authors declare that they have no conflict of interest.

\subsection*{Ethical approval}
The data used in this study were collected as part of a previous project approved by the Social and Societal Ethics Committee of KU Leuven (approval number: G-2023-7257). The present study is based on a secondary analysis of anonymized data.

\subsection*{Consent to participate}
Informed consent was obtained from all participants in the original study.

\subsection*{Consent for publication}
Not applicable.

\subsection*{Availability of data and materials}
The data used in this study are available at: 
\url{https://github.com/yufeiwu1011/EAMs-for-Bounded-Continuous-Data}.

\subsection*{Code availability}
The code used to conduct the analyses in this study is available at: \url{https://github.com/yufeiwu1011/EAMs-for-Bounded-Continuous-Data}. Both GPU and CPU versions of the scripts are provided.

\subsection*{Authors' contributions}
\noindent\textbf{Yufei Wu}: Conceptualization, methodology, software, formal analysis, writing—original draft. 
\textbf{Tamás Szűcs}: Data curation, resources, writing—review and editing. 
\textbf{Agnes Moors}: Resources, supervision, writing—review and editing. 
\textbf{Francis Tuerlinckx}: Conceptualization, methodology, supervision, writing—review and editing.

\newpage
\bibliography{ref}

@article{ratcliff1978theory,
  title={A theory of memory retrieval.},
  author={Ratcliff, Roger},
  journal={Psychological review},
  volume={85},
  number={2},
  pages={59},
  year={1978},
  publisher={American Psychological Association}
}

@article{von_krause_mental_2022,
	title = {Mental speed is high until age 60 as revealed by analysis of over a million participants},
	volume = {6},
	issn = {2397-3374},
	url = {https://www.nature.com/articles/s41562-021-01282-7},
	doi = {10.1038/s41562-021-01282-7},
	language = {en},
	number = {5},
	urldate = {2023-11-14},
	journal = {Nature Human Behaviour},
	author = {Von Krause, Mischa and Radev, Stefan T. and Voss, Andreas},
	month = feb,
	year = {2022},
	pages = {700--708},
}

@inproceedings{lipmanflow,
  title={Flow Matching for Generative Modeling},
  author={Lipman, Yaron and Chen, Ricky TQ and Ben-Hamu, Heli and Nickel, Maximilian and Le, Matthew},
  year={2022},
  booktitle={The Eleventh International Conference on Learning Representations}
}

@article{ratcliff_modeling_1998,
	title = {Modeling {Response} {Times} for {Two}-{Choice} {Decisions}},
	volume = {9},
	copyright = {http://journals.sagepub.com/page/policies/text-and-data-mining-license},
	issn = {0956-7976, 1467-9280},
	url = {http://journals.sagepub.com/doi/10.1111/1467-9280.00067},
	doi = {10.1111/1467-9280.00067},
	language = {en},
	number = {5},
	urldate = {2024-08-27},
	journal = {Psychological Science},
	author = {Ratcliff, Roger and Rouder, Jeffrey N.},
	month = sep,
	year = {1998},
	pages = {347--356},
}

@article{evans2020evidence,
  title={Evidence accumulation models: Current limitations and future directions},
  author={Evans, Nathan J and Wagenmakers, Eric-Jan and others},
  journal={The Quantitative Methods for Psychology},
  volume={16},
  number={2},
  pages={73--90},
  year={2020}
}

@article{gift1989visual,
  title={Visual analogue scales: measurement of subjective phenomena},
  author={Gift, Audrey G},
  journal={Nursing research},
  volume={38},
  number={5},
  pages={286--287},
  year={1989},
  publisher={LWW}
}

@article{mauss2009measures,
  title={Measures of emotion: A review},
  author={Mauss, Iris B and Robinson, Michael D},
  journal={Cognition and emotion},
  volume={23},
  number={2},
  pages={209--237},
  year={2009},
  publisher={Taylor \& Francis}
}

@book{clark1995constructing,
  title={Constructing validity: Basic issues in objective scale development.},
  author={Clark, Lee Anna and Watson, David},
  year={1995},
  publisher={American Psychological Association}
}

@article{smith2016diffusion,
  title={Diffusion theory of decision making in continuous report.},
  author={Smith, Philip L},
  journal={Psychological Review},
  volume={123},
  number={4},
  pages={425},
  year={2016},
  publisher={American Psychological Association}
}

@article{ratcliff2018decision,
  title={Decision making on spatially continuous scales.},
  author={Ratcliff, Roger},
  journal={Psychological review},
  volume={125},
  number={6},
  pages={888},
  year={2018},
  publisher={American Psychological Association}
}

@article{ratcliff2024using,
  title={Using diffusion models for symbolic numeracy tasks to examine aging effects.},
  author={Ratcliff, Roger and McKoon, Gail},
  journal={Journal of Experimental Psychology: Learning, Memory, and Cognition},
  volume={50},
  number={9},
  pages={1385},
  year={2024},
  publisher={American Psychological Association}
}

@article{berkhof2000posterior,
  title={Posterior predictive checks: Principles and discussion},
  author={Berkhof, Johannes and Van Mechelen, Iven and Hoijtink, Herbert},
  journal={Computational Statistics},
  volume={15},
  number={3},
  pages={337--354},
  year={2000},
  publisher={Springer}
}

@article{brown2008integrated,
  title={An integrated model of choices and response times in absolute identification.},
  author={Brown, Scott D and Marley, AAJ and Donkin, Christopher and Heathcote, Andrew},
  journal={Psychological review},
  volume={115},
  number={2},
  pages={396},
  year={2008},
  publisher={American Psychological Association}
}

@article{brown2008simplest,
  title={The simplest complete model of choice response time: Linear ballistic accumulation},
  author={Brown, Scott D and Heathcote, Andrew},
  journal={Cognitive psychology},
  volume={57},
  number={3},
  pages={153--178},
  year={2008},
  publisher={Elsevier}
}

@book{lord2014introduction,
  title={An introduction to computational stochastic PDEs},
  author={Lord, Gabriel J and Powell, Catherine E and Shardlow, Tony},
  volume={50},
  year={2014},
  publisher={Cambridge University Press}
}

@article{teoh_framing_2023,
    title = {Framing {Subjective} {Emotion} {Reports} as {Dynamic} {Affective} {Decisions}},
    volume = {4},
    issn = {2662-2041, 2662-205X},
    url = {https://link.springer.com/10.1007/s42761-023-00197-y},
    doi = {10.1007/s42761-023-00197-y},
    language = {en},
    number = {3},
    urldate = {2024-06-28},
    journal = {Affective Science},
    author = {Teoh, Yi Yang and Cunningham, William A. and Hutcherson, Cendri A.},
    month = sep,
    year = {2023},
    pages = {522--528},
}

@article{givon_are_2023,
    title = {Are women truly “more emotional” than men? {Sex} differences in an indirect model-based measure of emotional feelings},
    volume = {42},
    issn = {1936-4733},
    shorttitle = {Are women truly “more emotional” than men?},
    url = {https://doi.org/10.1007/s12144-022-04227-z},
    doi = {10.1007/s12144-022-04227-z},
    language = {en},
    number = {36},
    urldate = {2025-01-30},
    journal = {Current Psychology},
    author = {Givon, Ella and Berkovich, Rotem and Oz-Cohen, Elad and Rubinstein, Kim and Singer-Landau, Ella and Udelsman-Danieli, Gal and Meiran, Nachshon},
    month = dec,
    year = {2023},
    pages = {32469--32482},
}

@inproceedings{lee2019set,
  title={Set transformer: A framework for attention-based permutation-invariant neural networks},
  author={Lee, Juho and Lee, Yoonho and Kim, Jungtaek and Kosiorek, Adam and Choi, Seungjin and Teh, Yee Whye},
  booktitle={International conference on machine learning},
  pages={3744--3753},
  year={2019},
  organization={PMLR}
}

@article{zammit2024neural,
  title={Neural Methods for Amortized Inference},
  author={Zammit-Mangion, Andrew and Sainsbury-Dale, Matthew and Huser, Rapha{\"e}l},
  journal={Annual Review of Statistics and Its Application},
  volume={12},
  year={2024},
  publisher={Annual Reviews}
}

@misc{radev_bayesflow_2020,
    title = {{BayesFlow}: {Learning} complex stochastic models with invertible neural networks},
    shorttitle = {{BayesFlow}},
    url = {https://arxiv.org/abs/2003.06281},
    doi = {10.48550/ARXIV.2003.06281},
    howpublished = {arXiv:2003.06281},
    author = {Radev, Stefan T. and Mertens, Ulf K. and Voss, Andreas and Ardizzone, Lynton and K{\"o}the, Ullrich},
    year = {2020},
}

@book{rogers2000diffusions,
  title={Diffusions, Markov processes, and martingales},
  author={Rogers, L Chris G and Williams, David},
  volume={2},
  year={2000},
  publisher={Cambridge university press}
}

@article{radev2021amortized,
  title={Amortized bayesian model comparison with evidential deep learning},
  author={Radev, Stefan T and D’Alessandro, Marco and Mertens, Ulf K and Voss, Andreas and Koethe, Ullrich and Buerkner, Paul-Christian},
  journal={IEEE Transactions on Neural Networks and Learning Systems},
  volume={34},
  number={8},
  pages={4903--4917},
  year={2021},
  publisher={IEEE}
}

@article{elsemuller2023sensitivity,
  title={Sensitivity-aware amortized bayesian inference},
  author={Elsem{\"u}ller, Lasse and Olischl{\"a}ger, Hans and Schmitt, Marvin and B{\"u}rkner, Paul-Christian and K{\"o}the, Ullrich and Radev, Stefan T},
  journal={arXiv preprint arXiv:2310.11122},
  year={2023}
}

@article{cannon2022investigating,
  title={Investigating the impact of model misspecification in neural simulation-based inference},
  author={Cannon, Patrick and Ward, Daniel and Schmon, Sebastian M},
  journal={arXiv preprint arXiv:2209.01845},
  year={2022}
}

@article{elsemuller2024deep,
  title={A deep learning method for comparing Bayesian hierarchical models.},
  author={Elsem{\"u}ller, Lasse and Schnuerch, Martin and B{\"u}rkner, Paul-Christian and Radev, Stefan T},
  journal={Psychological Methods},
  year={2024},
  publisher={American Psychological Association}
}

@article{charness2013experimental,
  title={Experimental methods: Eliciting risk preferences},
  author={Charness, Gary and Gneezy, Uri and Imas, Alex},
  journal={Journal of economic behavior \& organization},
  volume={87},
  pages={43--51},
  year={2013},
  publisher={Elsevier}
}

@article{kleitman2007self,
  title={Self-confidence and metacognitive processes},
  author={Kleitman, Sabina and Stankov, Lazar},
  journal={Learning and individual differences},
  volume={17},
  number={2},
  pages={161--173},
  year={2007},
  publisher={Elsevier}
}

@article{trueblood2014multiattribute,
  title={The multiattribute linear ballistic accumulator model of context effects in multialternative choice.},
  author={Trueblood, Jennifer S and Brown, Scott D and Heathcote, Andrew},
  journal={Psychological review},
  volume={121},
  number={2},
  pages={179},
  year={2014},
  publisher={American Psychological Association}
}

@article{mulder_bias_2012,
	title = {Bias in the {Brain}: {A} {Diffusion} {Model} {Analysis} of {Prior} {Probability} and {Potential} {Payoff}},
	volume = {32},
	issn = {0270-6474, 1529-2401},
	shorttitle = {Bias in the {Brain}},
	url = {https://www.jneurosci.org/lookup/doi/10.1523/JNEUROSCI.4156-11.2012},
	doi = {10.1523/JNEUROSCI.4156-11.2012},
	language = {en},
	number = {7},
	urldate = {2024-03-08},
	journal = {The Journal of Neuroscience},
	author = {Mulder, Martijn J. and Wagenmakers, Eric-Jan and Ratcliff, Roger and Boekel, Wouter and Forstmann, Birte U.},
	month = feb,
	year = {2012},
	pages = {2335--2343},
}

@article{ratcliff_diffusion_2004,
	title = {A {Diffusion} {Model} {Account} of the {Lexical} {Decision} {Task}.},
	volume = {111},
	issn = {1939-1471, 0033-295X},
	url = {https://doi.apa.org/doi/10.1037/0033-295X.111.1.159},
	doi = {10.1037/0033-295X.111.1.159},
	language = {en},
	number = {1},
	urldate = {2024-10-28},
	journal = {Psychological Review},
	author = {Ratcliff, Roger and Gomez, Pablo and McKoon, Gail},
	year = {2004},
	pages = {159--182},
}

@unpublished{szucs_wu_tuerlinckx_moors2026,
  title = {A systematic examination of the determinants of affect in a drift-diffusion framework},
  author = {Szűcs, Tamás and Wu, Yufei and Tuerlinckx, Francis and Moors, Agnes},
  note = {Unpublished manuscript},
  year = {2026}
}

@article{wu2026testing,
  title={Testing and improving the robustness of amortized bayesian inference for cognitive models.},
  author={Wu, Yufei and Radev, Stefan T and Tuerlinckx, Francis},
  journal={Psychological Methods},
  year={2026},
  publisher={American Psychological Association}
}

@book{gelman2013bayesian,
  title     = {Bayesian Data Analysis},
  author    = {Gelman, Andrew and Carlin, John B. and Stern, Hal S. and Dunson, David B. and Vehtari, Aki and Rubin, Donald B.},
  edition   = {3},
  year      = {2013},
  publisher = {CRC Press}
}

@article{kvam2023unified,
  title={A unified theory of discrete and continuous responding.},
  author={Kvam, Peter D and Marley, AAJ and Heathcote, Andrew},
  journal={Psychological Review},
  volume={130},
  number={2},
  pages={368},
  year={2023},
  publisher={American Psychological Association}
}

@book{ekkekakis2013measurement,
  title={The measurement of affect, mood, and emotion: A guide for health-behavioral research},
  author={Ekkekakis, Panteleimon},
  year={2013},
  publisher={Cambridge University Press}
}

@article{molenaar2022zero,
  title={Zero and one inflated item response theory models for bounded continuous data},
  author={Molenaar, Dylan and C{\'u}ri, Mariana and Baz{\'a}n, Jorge L},
  journal={Journal of Educational and Behavioral Statistics},
  volume={47},
  number={6},
  pages={693--735},
  year={2022},
  publisher={Sage Publications Sage CA: Los Angeles, CA}
}

@article{sailynoja2022graphical,
  title={Graphical test for discrete uniformity and its applications in goodness-of-fit evaluation and multiple sample comparison},
  author={S{\"a}ilynoja, Teemu and B{\"u}rkner, Paul-Christian and Vehtari, Aki},
  journal={Statistics and Computing},
  volume={32},
  number={2},
  pages={32},
  year={2022},
  publisher={Springer}
}

\newpage

\begin{appendices}

\section*{Appendix}

\setcounter{table}{0}
\renewcommand{\thetable}{A\arabic{table}}
\setcounter{figure}{0}
\renewcommand{\thefigure}{A\arabic{figure}}
\setcounter{equation}{0}
\renewcommand{\theequation}{A\arabic{equation}}

\subsection*{The specular reflection in HCDM}
If a step $\Delta\boldsymbol{s}_t = (\Delta x_{1t}, \Delta x_{2t})$ would cross the boundary ($x_{2t} + \Delta x_{2t} < 0$), the step is decomposed into a part reaching the boundary and a reflective remainder. The part that reaches the boundary is $\boldsymbol{s}_b = \boldsymbol{s}_t + \tau\,\Delta\boldsymbol{s}_t$, with $\tau = \frac{-x_{2t}}{\Delta x_{2t}}$. The remaining step $\boldsymbol{s}_{\text{sem}} = (1-\tau) \Delta\boldsymbol{s}_t = (s_1, s_2)$ is then reflected, denoted as $\textbf{r}^{\text{ref}} = (s_1, -s_2)$. This yields the post-reflection position:
\begin{equation}
    \boldsymbol{s}_{t+1} = \boldsymbol{s}_b + \mathbf{r}^{\,\mathrm{ref}}.
    \label{eq:reflection}
\end{equation}
This procedure produces a mirror-like bounce at diameter $x_2=0$. 

\subsection*{ABI workflow}

\begin{figure}[!ht]
\centering
\includegraphics[width=1\linewidth]{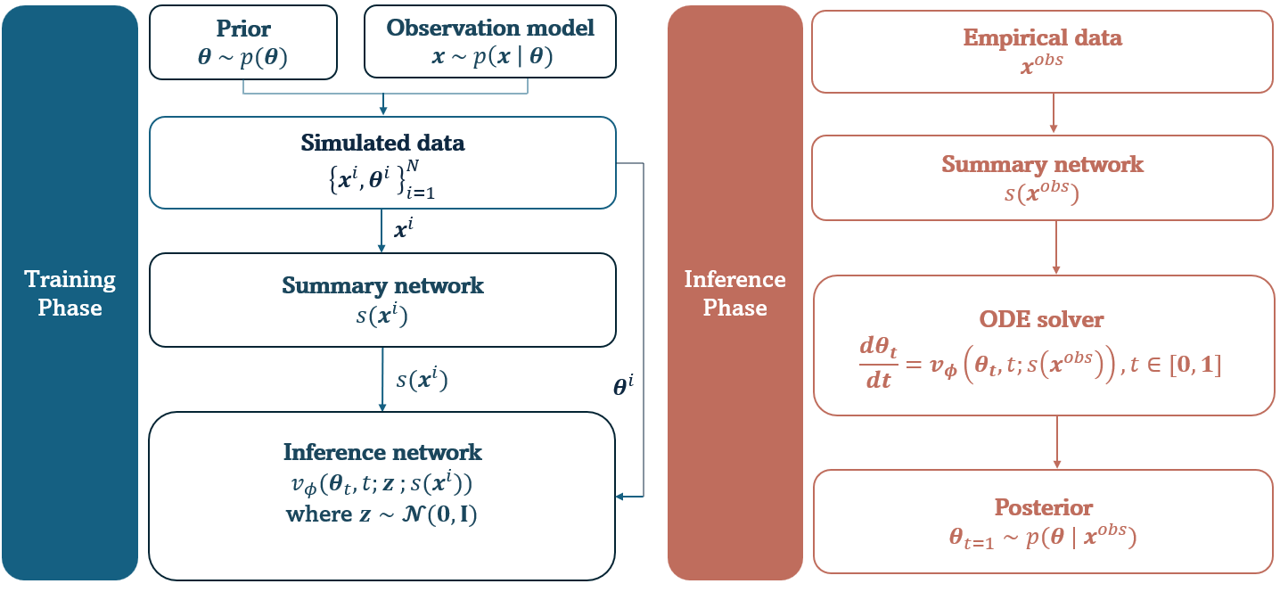}
\caption{\label{fig:BayesFlowworkflow} The basic workflow of conditional flow matching in ABI for posterior estimation. Parameters and data are simulated from a prior and an observation model. These simulations are used as training data for the summary and inference networks that jointly learn a vector field that transforms random noise to posterior samples from $t=0$ to $t=1$. Once trained, the networks can instantly sample from the posterior for any observed data by an ODE solver.}
\end{figure}

A basic workflow is depicted in Figure~\ref{fig:BayesFlowworkflow}. The first step is to generate synthetic data as training data. One first defines an observational model $\bx \sim p(\bx \mid \btheta)$ with a prior $\btheta \sim p(\btheta)$ and, where $\btheta \in \mathbb{R}^D$. Parameters are simulated from the prior and passed to the observational model to generate training data. The simulated parameters and data $\{\bx^i, \btheta^i\}_{i=1}^I$ are used to train two neural networks: a summary and an inference network. The summary network transforms each data set $\bx^i$ into fixed-size \textit{approximately sufficient} summary statistics $s(\bx^i)$, where $s(\cdot)$ represents the transformation applied by the summary network. The inference network approximates complicated posterior distributions via algorithms such as conditional flow matching \citep[CFM,][]{lipmanflow}, one of the most expressive architectures that is currently available. With CFM, the inference network learns a time-dependent vector field $v_t$ that can map a predefined simple base distribution $\boldsymbol{z}$ (e.g., a spherical Gaussian) to a complex target distribution, where $t \in [0, 1]$:
\begin{equation*}
    \btheta \sim p(\btheta \mid \bx) \Longleftrightarrow \btheta = \psi_1 (\boldsymbol{z};\,s(\bx)) \quad \text{with} \quad \boldsymbol{z} \sim \mathcal{N}(\mathbf{0}, \mathbf{I}), \label{eq:flow_matching}
\end{equation*}
where $\psi_1$ is the flow map generated by integrating the conditional vector field $v_t$ from $t=0$ to $t = 1$.

Upon convergence, the vector field $v_\phi$ is learned, drawing samples from the approximate posterior $q(\btheta \mid s(\bx^{\text{obs}}))$ involves solving an ordinary differential equations (ODE). Given an observation $\bx^{\text{obs}}$, we sample an initial state from the base distribution, $\btheta_{t=0} = \boldsymbol{z} \sim \mathcal{N}(\mathbf{0}, \mathbf{I})$, and simulate the continuous-time dynamics governed by the learned vector field:
\begin{equation*}
    \frac{\mathrm{d}\btheta_t}{\mathrm{d}t} = v_\phi(\btheta_t, t; s(\bx^{\text{obs}})), \quad t \in [0, 1].
\end{equation*}
The final state at $t=1$, denoted as $\btheta_{t=1}$, constitutes a sample from the targeted posterior distribution. In practice, this integration is performed using a numerical ODE solver (e.g., the Euler method).

\subsection*{ABMC workflow}

  Using the method developed by \cite{radev2021amortized}, we consider the PMPs as probability parameters of probability simplex (because they sum to one) and we estimate the parameters of a Dirichlet distribution over this probability simplex. The Dirichlet distribution is defined as:
\begin{equation*}
    \text{Dir}( \bp \mid \balpha) = \frac{1}{B(\boldsymbol{\alpha})} \prod_{j=1}^{J} p_j^{\alpha_j - 1},
\end{equation*}
where $\bp$ is a probability simplex, defined as $\bp = \{ p(\mathcal{M}_1 \mid \bx_\text{obs}),..., p(\mathcal{M}_J \mid \bx_\text{obs}) \}$, and $B(\balpha)$ is the multivariate beta function. Note that in our case, the Dirichlet actually reduces to a Beta distribution (because there are only two models).

The parameter vector $\balpha$ summarizes the evidence over the models (outputs). It can be seen as a vector of concentration parameters $\balpha \in \mathbb{R}^J_+$ that determines the PDF of the Dirichlet distribution over $J$ probabilities that sum to 1. 
 
We can use the mean of the Dirichlet distribution, which is a vector of probabilities given by:
\begin{equation*}
    \mathbb{E}_{\bp \sim \text{Dir}(\balpha)}[\bp] = \balpha \frac{1}{\alpha_0}
\end{equation*}
where $\alpha_0 = \sum_{j=1}^J \alpha_j$, as an approximation to the posterior model probabilities. In fact, the neural network will output $\balpha$. 

It is important to note that this method implicitly favors simpler models. Data generated by a simpler model tend to be more homogeneous than data generated by a more complex competitor. As a result, during training, datasets that are plausible under both models are more likely to be generated by the simpler model than by the more complex one.

\begin{figure}[ht!]
    \centering
    \includegraphics[width=1\linewidth]{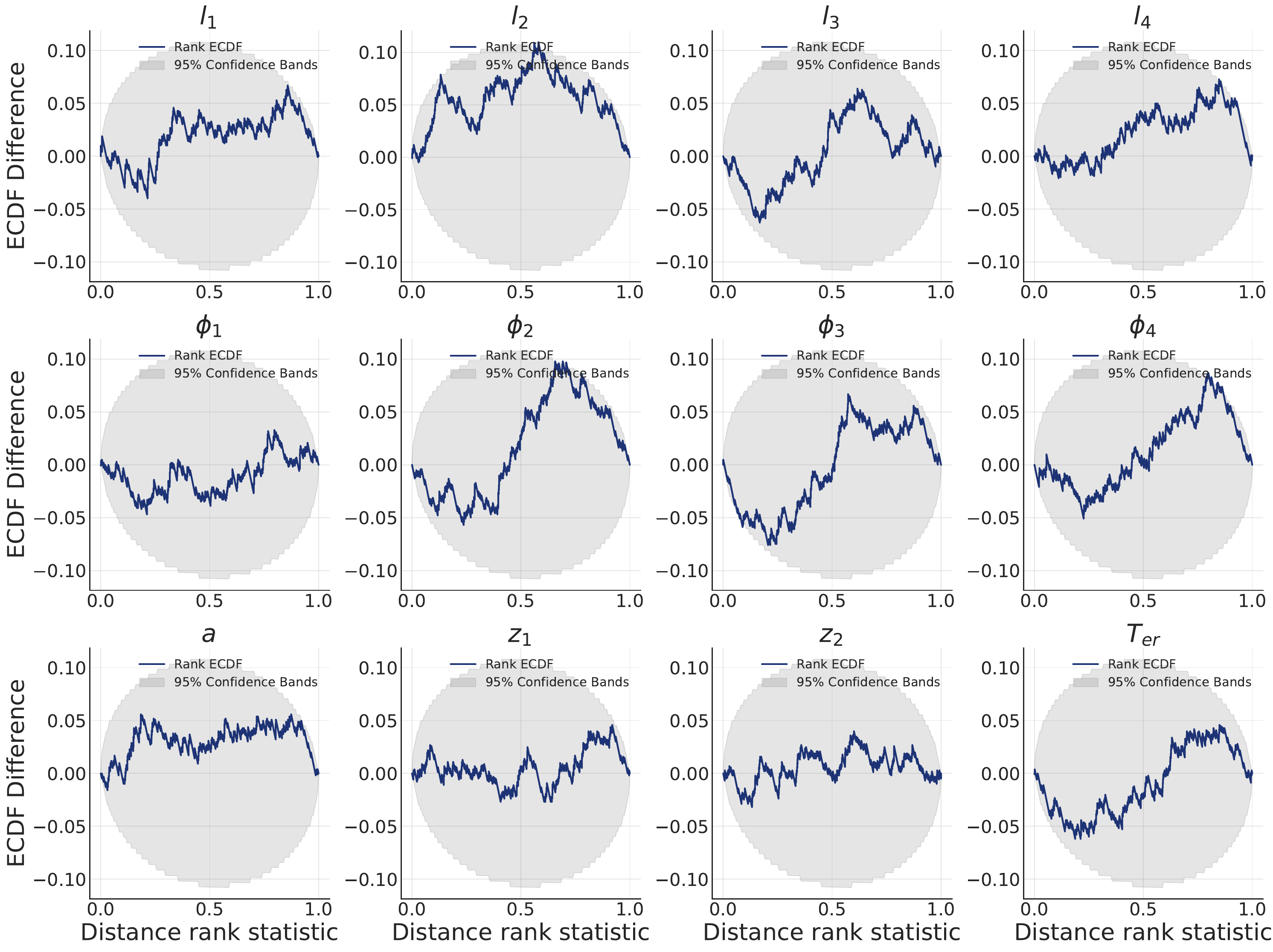}
    \caption{The empirical cumulative density function (ECDF) of rank statistics plotted against uniform ECDF in HCDM. If the posterior ranks of the prior draws distributed uniformly, the estimator yielded correct posteriors. For details of rank statistics, please refer to  Säilynoja et al. (\citeyear{sailynoja2022graphical}).}
    \label{fig:ecdf_hcdm}
\end{figure}

\begin{figure}[ht!]
    \centering
    \includegraphics[width=1\linewidth]{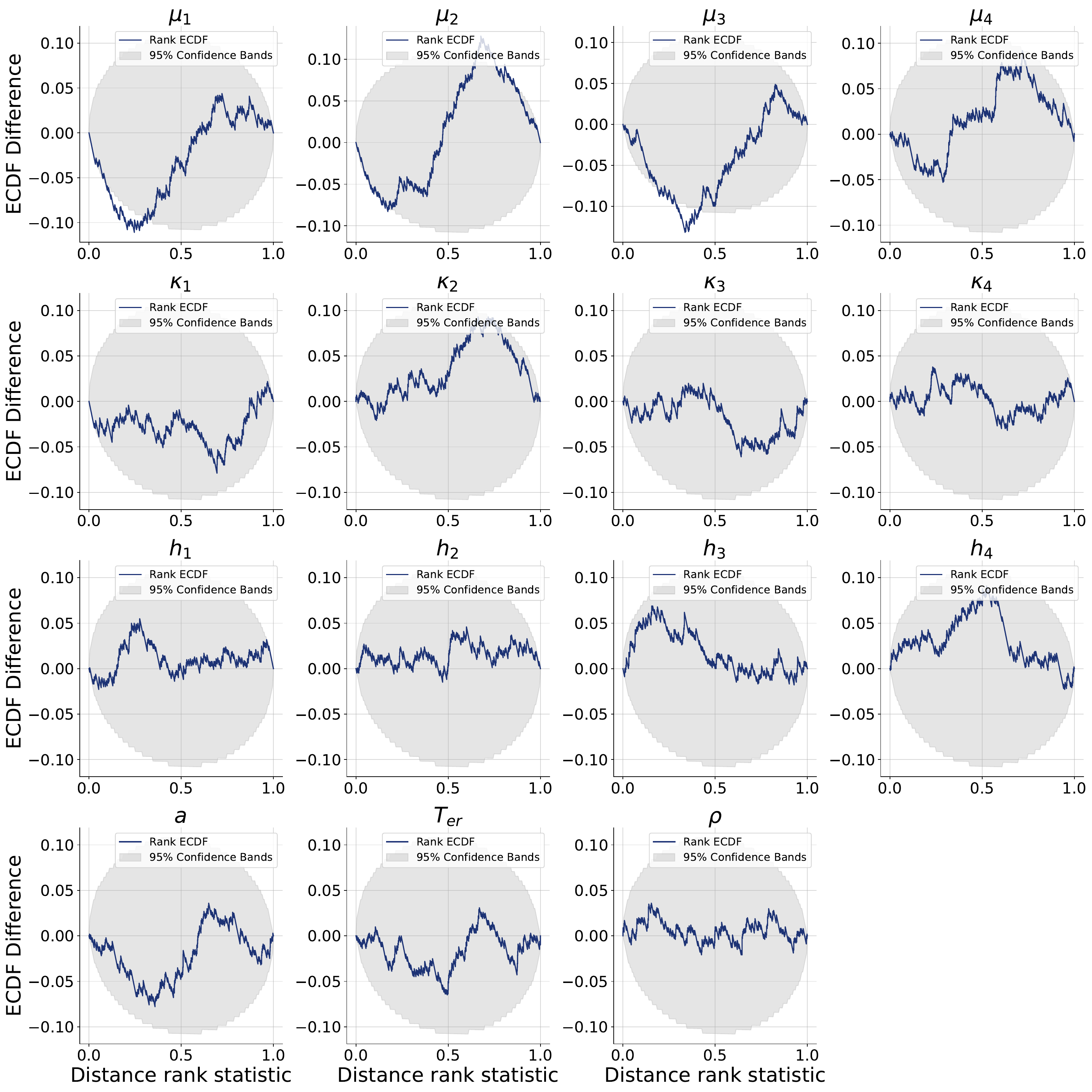}
    \caption{The empirical cumulative density function (ECDF) of rank statistics plotted against uniform ECDF in BDDM. If the posterior ranks of the prior draws distributed uniformly, the estimator yielded correct posteriors. For details of rank statistics, please refer to  Säilynoja et al. (\citeyear{sailynoja2022graphical}).}
    \label{fig:ecdf_bddm}
\end{figure}

\begin{figure}[!ht] 
 \centering
 \includegraphics[width=1\linewidth]{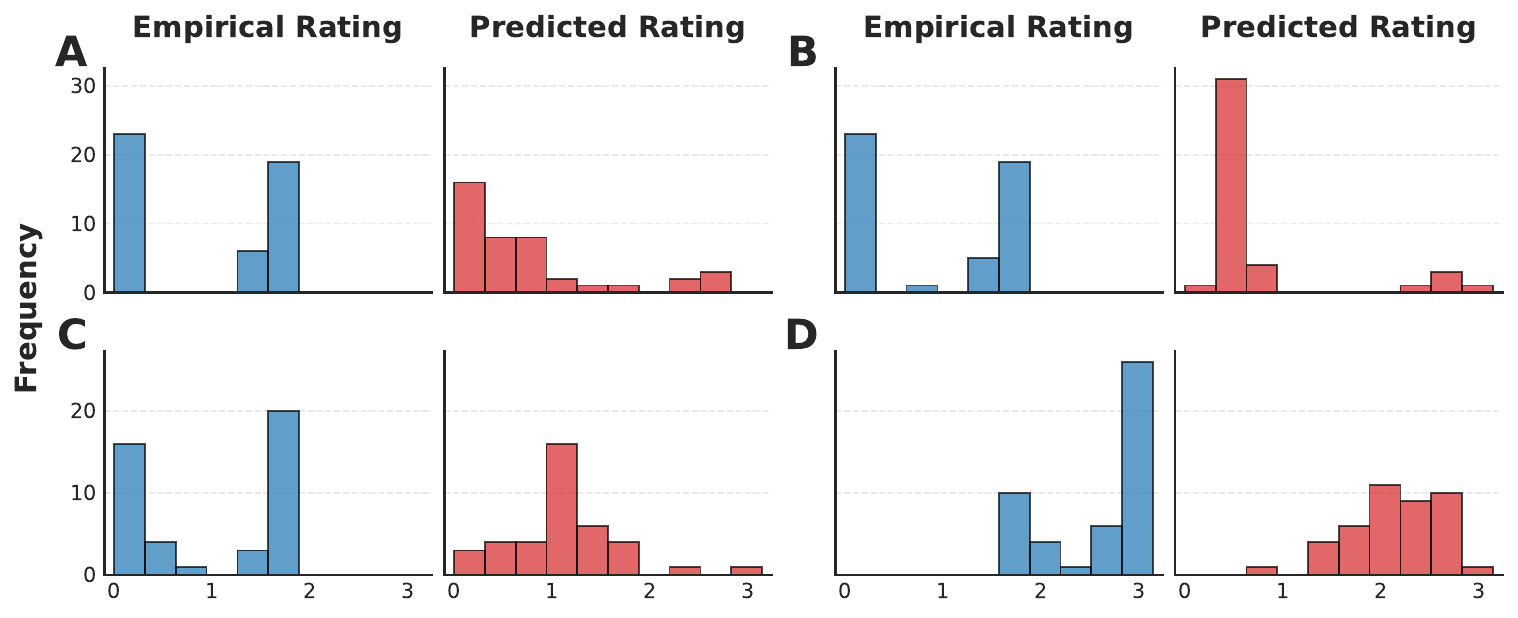}
 \caption{Inaccurate estimation of the 0.5 quantile of rating using the BDDM. Panels A-D show four different participants. The empirical and predicted ratings shown are specifically from the condition where the model failed to accurately predict its 0.5 quantile when the observed quantile is around the midpoint of the scale. } 
 \label{fig:BDDMprobrating2}
\end{figure}

\begin{figure}[ht!]
    \centering
    \includegraphics[width=1\linewidth]{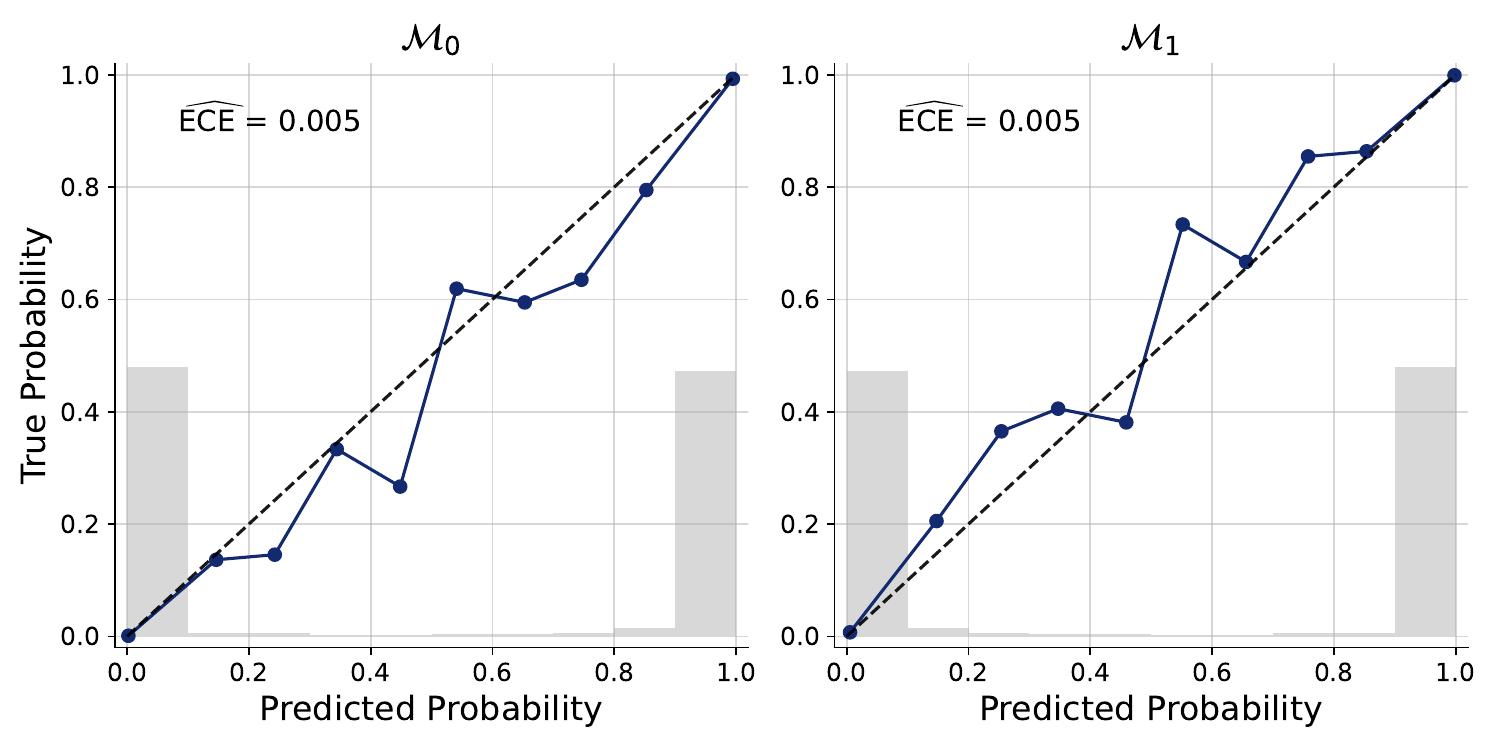}
    \caption{Calibration curve for a PMP network. We simulated 10,000 datasets and grouped the predicted PMPs into 10 bins (x-axis). For each bin, we computed the empirical fraction of datasets generated by the model (y-axis). The black dashed line indicates perfect calibration. $\hat{\mathrm{ECE}}$, the expected calibration error, was computed as the average absolute deviation between predicted and empirical probabilities across bins.}
    \label{fig:mccalibration}
\end{figure}

\begin{figure}[ht!]
    \centering
    \includegraphics[width=0.5\linewidth]{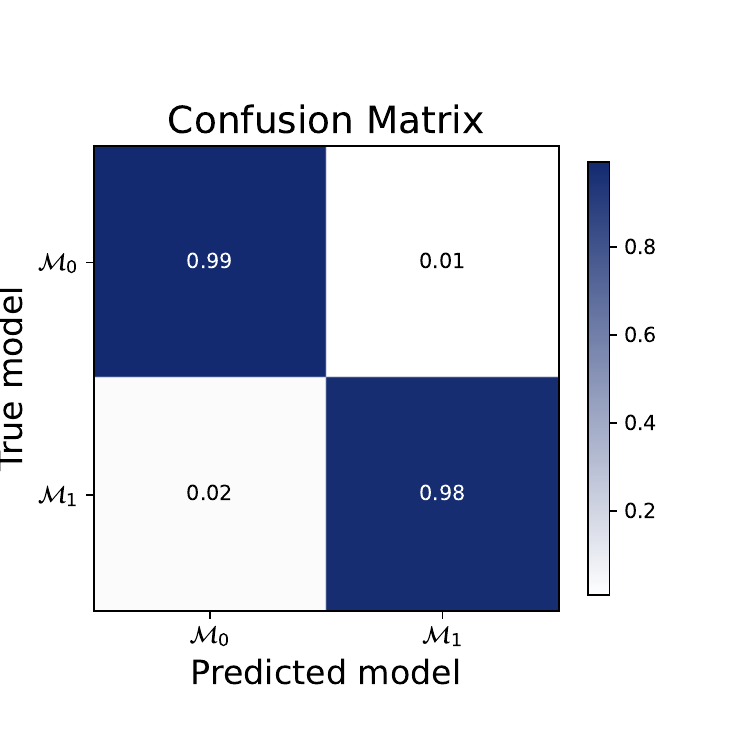}
    \caption{Confusion matrix of a PMP network. $\mathcal{M}_0$ stands for the HCDM, and $\mathcal{M}_1$ represents the BDDM.}
    \label{fig:confusionmatrix}
\end{figure}
\end{appendices}
\end{document}